\documentclass[12pt]{article}

\usepackage{graphicx}
\usepackage{epstopdf}
\usepackage{amsmath}
\usepackage{amsfonts}
\usepackage{amssymb}
\usepackage{color}

\def\vep{\varepsilon}
\def\mpl{M_{pl}}
\def\ta{{\tilde a}}
\def\tf{{\tilde f}}
\def\tA{{\tilde A}}

\def\ts{{\tilde s}}
\def\bg{{\bar g}}
\def\bA{{\bar A}}
\def\ba{{\bar a}}
\def\baf{{\bar f}}

\def\bv{{\bar v}}

\def\beq{\begin{eqnarray}}
\def\eeq{\end{eqnarray}}

\newcommand{\comment}[1]{}

\begin{document}

\begin{titlepage}

\topmargin 1.5 cm
\centerline{ \LARGE A Proof Of Ghost Freedom}

\vskip 0.6 cm
\centerline{\LARGE In de Rham-Gabadadze-Tolley Massive Gravity}

\vskip 2 cm
\centerline{\Large Mehrdad Mirbabayi}
\vskip 0.5 cm
\centerline{\small\em CCPP, Physics Department, New York University, 4 Washington place, New York, NY}


\vskip 2.5 cm
\begin{abstract}

\noindent\normalsize We identify different helicity degrees of freedom of Fierz-Paulian massive gravity around generic backgrounds. We show that the two-parameter family proposed by de Rham, Gabadadze, and Tolley always propagates five degrees of freedom and therefore is free from the Boulware-Deser ghost. The analysis has a number of byproducts, among which (a) it shows how the original decoupling limit construction ensures ghost freedom of the full theory, (b) it reveals an enhanced symmetry of the theory around linearized backgrounds, and (c) it allows us to give an algorithm for finding dispersion relations. The proof naturally extends to generalizations of the theory with a reference metric different from Minkowski.

\end{abstract}

\end{titlepage}

\parskip 0.23 cm

\section{Introduction and Summary}

The unique Lorentz-invariant and ghost-free linearized theory of massive gravity was first formulated by Fierz and Pauli \cite{FP}. As such, the theory propagates the right number of five degrees of freedom, pertinent to a massive spin-2 particle, when considered sufficiently close to Minkowski background. However, through an analysis of constraints, Boulware and Deser showed that an extra negative-energy degree of freedom arises at nonlinear level \cite{BD} (the so-called Boulware-Deser ghost). 

The analysis of Boulware and Deser, while being generically applicable to Fierz-Paulian massive gravity---the set of nonlinear completions of the Fierz-Pauli mass term (FP), overlooked the possibility of a clever choice of the potential at higher orders so as to avoid the sixth degree of freedom. Using a more tractable description of massive gravity based on St\"{u}eckelberg fields, proposed by Arkani-Hamed, Georgi, and Schwartz (AGS) \cite{AGS}, de Rham and Gabadadze made a promising step in that direction \cite{giga1}. They constructed a two-parameter subfamily of FP that was free from the Boulware-Deser ghost in a special, yet nontrivial limit.

The theory, first formulated as an infinite series of higher order corrections to quadratic Fierz-Pauli action, was later nicely resummed by de Rham, Gabadadze, and Tolley \cite{giga2}. This made possible an extensive investigation of different classical solutions of the theory and comparison with General Relativity (GR).\footnote{For some examples see \cite{Koyama,theo,stars,Volkov,giga_cosmo,IPMU,Mohseni,KoyamaSA,Berezhiani,Mukoyama}. For a review see \cite{Kurt}.} We henceforth refer to this two-parameter subfamily of FP as dRGT. 

dRGT certainly has many good features compared to generic FP. First of all, and by construction, it is free from the Boulware-Deser ghost in the decoupling limit provided that there is no background vector field \cite{giga1}. Unlike generic FP, in which the asymptotically flat spherically symmetric solution is not unique (essentially because the order of equations of motion is higher than 2) \cite{Babichev}, dRGT has a unique solution \cite{stars}. Moreover, the homogeneous and isotropic solution is stable (actually static) \cite{giga_cosmo}, in contrast to the unstable solutions in generic FP \cite{giga_andrei}. These all suggest that the theory should be completely free from the sixth degree of freedom.

An elegant proof of this was first found by Hassan and Rosen \cite{Rachel}. Using a special representation of dRGT \cite{Rachel_resum}, they showed, in the canonical formalism, that after a redefinition of the shift vector the lapse remains a Lagrange multiplier and provides necessary constraints to remove the Boulware-Deser ghost (for complementary discussions see \cite{Kluson} and \cite{Rachel_confirm}). The proof was later extended to generalizations of dRGT with an arbitrary, instead of Minkowski, reference metric \cite{Rachel_ref} and bigravity models \cite{Rachel_bigrav}.\footnote{The phenomenology of the resulting bigravity models have been considered in \cite{VolkovII,Strauss,LuigiII,Pilo}.}

In this paper to prove the absence of the sixth degree of freedom we take another approach which is based on the following notion. Although the fluctuations of FP close to Minkowski background are stable, the way in which Boulware and Deser's non-linear ghost manifests itself is that the spectrum of fluctuations around other backgrounds generically contains an extra ghost-like mode. Therefore, if one shows that in a particular member of FP the spectrum of perturbations around all backgrounds contains five degrees of freedom, that theory is free from the Boulware-Deser ghost. Obviously if the analysis can be carried out only in a continuous region of the configuration space, as is the case here (see Sec. \ref{sec:ex}), the conclusion holds only in that region. So, our strategy would be to identify different helicity degrees of freedom around generic backgrounds and show that in dRGT there are (almost) always five modes. The proof easily generalizes to theories with arbitrary reference metrics. 

To perform the analysis we use the fact that any Lorentz-invariant theory of massive gravity can be formulated as GR plus a theory of four scalar fields $\phi^A$, $A=0,1,2,3$, with an internal Lorentz symmetry. The essence of the proof is to align the field-space basis of the scalars $\phi^A$, with the coordinate basis of the given background. As long as this alignment is possible one of the scalar fields can be shown to be nondynamical in dRGT, while generically all of them are dynamical in FP. We will see that the alignment is possible unless the unitary gauge metric is too far from Minkowski.

It is worth noting that the above scalar-tensor formulation of dRGT greatly simplifies in two space-time dimensions and the full nonlinear theory can be shown to propagate zero degrees of freedom \cite{giga_stu,Kluson_2d}. We were unable to generalize those methods to higher dimensional versions; instead, we concentrated on the behavior of the theory expanded around different backgrounds.

As a byproduct, our approach illuminates a number of (so far) less explored aspects of FP in general, and dRGT in particular:

\begin{itemize}
\renewcommand {\labelitemi}{$\bullet$}

\item It allows us to understand when the rather controversial St\"{u}eckelberg field that was introduced by AGS to describe the helicity-0 mode fails to do so. Our formalism provides the alternative definition that is always applicable (let us call this field $\pi$).

\item It becomes clear how the original criterion imposed by de Rham and Gabadadze, namely, the absence of self-interactions of $\pi$, directly results in the absence of the Boulware-Deser ghost in the full theory.

\item It enables us to give a practical prescription to obtain dispersion relations of vector and scalar modes around generic backgrounds.

\item It follows from our analysis that in dRGT, if one freezes the metric to be flat, the aforementioned theory of four scalar fields becomes degenerate on linearized backgrounds -- it propagates only two degrees of freedom. Therefore, mixing with gravity is necessary to give dynamics to the third mode and the cutoff of the theory becomes of order $\Lambda_3=(\mpl m^2)^{1/3}$.

\item Once nonlinearities with a length scale $l$ are considered, for $l\ll m^{-1}$ gravity can be decoupled and the cutoff increases to $\Lambda_l=(\mpl m/l)^{1/3}$.

\end{itemize}

We emphasize that despite the absence of the sixth mode in dRGT, the two vector and one scalar degrees of freedom may become superluminal \cite{Andrei,giga_suplum} or ghostlike \cite{giga4,KoyamaSA} around some backgrounds. We hope our approach assists similar analyses around a larger class of solutions.

The organization of the paper is as follows. In Sec. \ref{sec:pre}, we review FP and dRGT with an emphasis on their relevant properties for the rest of the discussion. In Sec. \ref{sec:lin}, we study the behavior of the theory around the important class of linearized backgrounds. The unique feature of dRGT will become evident at this level. In Sec. \ref{sec:nl}, we extend the analysis to nonlinear backgrounds to complete the proof of ghost freedom. And finally, in Sec. \ref{sec:disp} we provide a recipe to find dispersion relations of vector and scalar modes. Given the contribution of nonlinearities to the dynamics of the scalar mode, we determine the upgraded cutoff of the theory.

\section{\label{sec:pre}Preliminaries}

The following subsections are meant to provide a concise but relatively self-contained review of (a) Fierz-Pauli massive gravity from a modern point of view, and (b) the built-in features of de Rham-Gabadadze-Tolley theory. For the most part, we follow discussions of \cite{AGS} and \cite{giga1}, with the addition of a few complementary comments, calculations, and corollaries. 

\subsection{\label{sec:scalar}Massive Gravity as a Theory of Four Scalar Fields}

Let us start with the Fierz-Pauli mass term for gravity
\beq
-\frac{1}{4}\mpl^2m^2 \int d^4x(h_{\mu\nu}^2-h^2)\,,
\label{FP}
\eeq
where $m$ is the graviton mass and $h_{\mu\nu}=g_{\mu\nu}-\eta_{\mu\nu}$ is the deviation of the metric from Minkowski. The contraction of indices can be done using either $\eta^{\mu\nu}$ or $g^{\mu\nu}$ since the theory is specified only up to quadratic order in $h_{\mu\nu}$. 

The action \eqref{FP} appears to break reparametrization invariance of GR. However the invariance can be easily restored by introducing four scalar fields $\phi^A$, $A=0,1,2,3$, also known as St\"{u}eckelberg fields \cite{AGS} (see also \cite{Siegel}). Defining the covariant tensor
\beq
\label{Hmn}
H_{\mu\nu}=g_{\mu\nu}-\eta_{AB}\partial_\mu\phi^A\partial_\nu\phi^B\,,
\eeq
a reparametrization invariant version of \eqref{FP} can be written as 
\beq
\label{FPH}
-\frac{\mpl^2m^2}{4}\int d^4x\sqrt{-g}g^{\mu\nu}g^{\rho\sigma}
(H_{\mu\rho}H_{\nu\sigma}-H_{\mu\nu}H_{\rho\sigma})\,.
\eeq
One can now fix the unitary gauge by choosing $x^A=\phi^A$ to recover the original action \eqref{FP}. 

Given the ambiguity of the Fierz-Pauli action beyond quadratic order it is natural to consider adding higher order corrections in $H_{\mu\nu}$ to \eqref{FPH} to form a family of Fierz-Paulian theories of massive gravity, which we denote in short by FP. The action of the most general FP can then be written as 
\beq
S&=&S_{EH}+S_{FP}\,,\\
\label{EH}
S_{EH}&=&\mpl^2 \int d^4x \sqrt{-g}R\,,\\
\label{FPV}
S_{FP}&=&-\frac{m^2 M_{pl}^2}{4}\int d^4x\sqrt{-g}~V(H^\mu_\nu)\,,
\eeq
where $V$ depends only on the invariants made out of $H^\mu_\nu=g^{\mu\sigma}H_{\sigma\nu}$, and, upon expansion around $H^\mu_\nu=0$, it reproduces \eqref{FPH} at leading order. 

Recalling the definition of $H_{\mu\nu}$ in \eqref{Hmn}, we see that a function of invariants made of $H^\mu_\nu$ effectively depends only on the invariants made out of the tensor $g^{\mu \sigma}\partial_ \sigma \phi^A \partial_ \nu \phi^B \eta _{AB}$. So we can write
\beq
\label{Vphi}
V=V(g^{\mu \sigma}\partial_ \sigma \phi^A \partial_ \nu \phi^B \eta _{AB})\,,
\eeq
which makes it manifest that any FP is equivalent to GR plus a theory of four scalar fields $\phi^A$. This is in fact valid for all theories of massive gravity; see \cite{Sergei} for non-FP examples. These scalar fields are noncanonical (have dimension of length), and are derivatively coupled (hence have a shift symmetry). In addition, the action remains invariant under a global internal Lorentz transformation
\beq
\phi^A\to \Lambda^A_B\phi^B,
\eeq
which is necessary to ensure the Lorentz invariance of the theory in the Higgsed phase $\phi^A=x^A$.

Thus we see that to covariantly define FP, four new scalar fields have been added to GR. Since the theory is now generally covariant the metric is expected to have only two tensor degrees of freedom. One therefore naively expects to get a theory with a total of six degrees of freedom, exceeding what is needed for a massive spin-2 graviton by one. 

However, as is well known, the special structure of the Fierz-Pauli mass term is such that this extra degree of freedom is absent in the linearized theory. To see this in the St\"{u}eckelberg formulation and also to set up a basis for the rest of the paper, we first introduce some useful notations and comment on them. 

\subsection{Some Notations}

Instead of fixing the unitary gauge let us write $\phi^A$, after changing its index to Greek, as
\beq
\label{A}
\phi^\mu=x^\mu-A^\mu\,,
\eeq
and $g_{\mu\nu}$ as
\beq
g_{\mu\nu}=\eta_{\mu\nu}+h_{\mu\nu},
\eeq
so that \eqref{Hmn} becomes
\beq
H_{\mu\nu}=h_{\mu\nu}+\partial_\mu A_\nu+\partial_\nu A_\mu
                                   -\partial_\mu A^\sigma\partial_\nu A_\sigma\,,
\eeq
where $A_\mu=\eta_{\mu\nu}A^\nu$. We can then formally expand the action \eqref{FPV} in terms of $A^\mu$ and $h_{\mu\nu}$, and the resulting action can be viewed as an interacting theory of a tensor $h_{\mu\nu}$, and a seemingly vector $A^\mu$ on flat space. To avoid carrying along the volume integrals we use Lagrangian density and decompose it into three pieces
\beq
\label{LFP}
-\frac{1}{4}\sqrt{-g}V(H^\mu_\nu)=\mathcal{L}_h+\mathcal{L}_{hA}+\mathcal{L}_A\,,
\eeq
a potential for $h_{\mu\nu}$, couplings between $h_{\mu\nu}$ and $A^\mu$, and a pure $A^\mu$ Lagrangian. We frequently refer to $\mathcal{L}_A$ as the pure scalar sector as it describes the fluctuations of the scalar fields $\phi^\mu$ on a frozen flat metric, $h_{\mu\nu}=0$.

Note that in \eqref{LFP} $A^\mu$ always appears with one derivative acting on it and $h_{\mu\nu}$ without derivative. All indices are raised and lowered by the Minkowski metric, and for the sake of simplicity we sacrificed the mass dimension so that $\mathcal{L}_h,\mathcal{L}_{hA}$, and $\mathcal{L}_A$ are dimensionless.

Note also that the notational twist from Latin to Greek indices in \eqref{A} does not, by itself, make $A^\mu$ a vector field, rather, the symmetries of the effective action around a given background determine how the fields regroup into representations of the Poincar\'{e} algebra. Around Minkowski background, where
\beq
h^{bg}_{\mu\nu}=\langle h_{\mu\nu}\rangle=0\,\quad\text{and}\quad A_{bg}^\mu=\langle A^\mu\rangle=0\,,
\eeq
a space-time Lorentz transformation $x'^\mu=\Lambda_\nu^\mu x^\nu$, if followed by the same Lorentz transformation in the field space $\phi'^\mu=\Lambda^\mu_\nu\phi^\nu$, gives an identical action for the Lorentz transformed field $A'^\mu=\Lambda^\mu_\nu A^\nu$. Therefore, we can call $A^\mu$ a vector field--the perturbations of the four scalar fields form the components of a vector field around Minkowski space. As will be thoroughly discussed, this is not necessarily the case around generic backgrounds and the use of Greek indices is just for convenience since we are expanding $g_{\mu\nu}$ around the Minkowski metric, the same metric as $\eta_{AB}$ which contracts internal indices. 

\subsection{\label{dofFP}Counting Degrees of Freedom of FP Close to Minkowski}

Now we can return to the analysis of Fierz and Pauli's special construct. If we look at the quadratic Lagrangian of $A_\mu$ and $h_{\mu\nu}$, that is, if we consider small fluctuations around $g_{\mu\nu}=\eta_{\mu\nu}$ and $\phi^\mu=x^\mu$, it is unambiguously dictated by the Fierz-Pauli term \eqref{FPH}. The pure scalar sector looks like
\beq
\label{FF}
\mathcal{L}_A^{(2)}=-\frac{1}{4}(2\partial_\mu A_\nu\partial_\mu A_\nu+2\partial_\mu A_\nu\partial_\nu A_\mu-4\partial_\mu A_\mu\partial_\nu A_\nu) =-\frac{1}{4} F_{\mu\nu}^2 +\text{total derivative.}
\eeq
This is the Maxwell theory that propagates two transverse vector degrees of freedom with helicity 1. Therefore in the pure scalar sector $\mathcal{L}_A$, the perturbations of the four scalar fields acquired an accidental $U(1)$ gauge symmetry, making two of them nondynamical. 

Of course $\mathcal{L}_A$ is not the whole story and what provides the dynamics of the fifth, helicity-0 degree of freedom is the mixing of $A^\mu$ with gravity:
\beq
\label{LhA2}
\mathcal{L}_{hA}^{(2)}=-\frac{1}{2}h^{\mu\nu}(\partial_\mu A_\nu+\partial_\nu A_\mu-2\eta_{\mu\nu}\partial_\sigma A^\sigma)\,.
\eeq
This term breaks the $U(1)$ gauge symmetry and makes another component of $A^\mu$ dynamical. 

To show this, we slightly modify Boulware and Deser's analysis of quadratic FP action. In the ADM formalism \cite{ADM}, the quadratic GR action can be put in the canonical form in which the perturbations of lapse and shift, $h_{00}$ and $h_{0i}$, are Lagrange multipliers: They do not carry any time derivative, which makes them nondynamical, and they appear linearly in the action. Thus the equations of motion for $h_{00}$ and $h_{0i}$ remove four out of six {\it a priori} degrees of freedom in $h_{ij}$ to give two transverse-traceless tensor modes. 

After the addition of the covariant Fierz-Pauli term \eqref{FPH}, the role of $h_{00}$ and $h_{0i}$ as Lagrange multipliers is preserved because of the general covariance of the theory. Moreover, the zeroth component of $A_\mu$ becomes nondynamical. This is because the only place that the time derivative of $A_0$ appears in quadratic FP is in $\mathcal{L}_{hA}^{(2)}$, where from \eqref{LhA2} we have 
\beq
-h_{ii}\dot A_0\,,
\eeq
with a summation over repeated indices. This can be integrated by parts to give a time derivative of an already dynamical field $h_{ij}$ and make $A_0$ nondynamical. Therefore, only three degrees of freedom are added to GR.

Therefore, our general strategy to count degrees of freedom around a given background will be to first look at the pure scalar sector $\mathcal{L}_A$ and see how many components of $A_\mu$ are dynamical there. If $\mathcal{L}_A^{(2)}$ was degenerate, that is, if one or two components remained nondynamical, then we investigate $\mathcal{L}_{hA}^{(2)}$. $\mathcal{L}_{hA}^{(2)}$ generally breaks the accidental $U(1)$ gauge symmetry that may arise in $\mathcal{L}_A^{(2)}$. However as was the case here, upon integration by parts it can conceivably leave one of the components of $A_\mu$ nondynamical without making $h_{00}$ and $h_{0i}$ dynamical. If so, we are ensured by the general covariance of the theory to have five degrees of freedom; $h_{00}$ and $h_{0i}$ remove four out of nine {\it a priori} degrees of freedom coming from six components of $h_{ij}$ and three components of $A_\mu$.

Let us confirm the above argument by an explicit calculation in the case of quadratic FP action which after separating time and space indices becomes
\beq
\label{SFP2}
S_{FP}^{(2)}&=&\mpl^2m^2\int d^4x \left[\frac{1}{2}F_{0i}^2-\frac{1}{4}F_{ij}^2
               -h_{00}\partial_iA_i
      +h_{0i}(\dot A_i+\partial_iA_0)\right.\nonumber\\
&&\left. -h_{ij}(\partial_iA_j-\delta_{ij}(-\dot A_0+\partial_iA_i))
  -\frac{1}{4}(h_{ij}^2-h_{ii}^2+2h_{00}h_{ii}-2h_{0i}^2)\right]\,.
\eeq
It is seen that unlike $h_{00}$, $h_{0i}$ is not manifestly a Lagrange multiplier. However, it becomes manifest once the nondynamical field $A_0$ is integrated out. Using the reparametrization invariance of the theory to fix the gauge 
\beq
\label{gauge}
h_{ii}=0,\qquad \partial_ih_{ij}=0\,,
\eeq
and also decomposing $h_{0i}$ and $A_i$ into transverse and longitudinal parts
\beq
\label{hT}
h_{0i}=h_{0i}^T+\partial_i\psi\,,\\
\label{AT}
A_i=A_i^T+\partial_i\varphi\,,
\eeq
with $\partial_ih_{0i}^T=0$ and $\partial_iA_i^T=0$, we get from the equation of motion for $A_0$ that
\beq
A_0=\dot \varphi-\psi\,.
\eeq
Substituting this expression for $A_0$ back in \eqref{SFP2} yields
\beq
S_{FP}^{(2)}=\mpl^2m^2\int d^4x \left[\frac{1}{2}{\dot {A_i^T}}^2
 +h_{0i}^T\dot A_i^T+\frac{1}{2}{h_{0i}^T}^2-\Delta\psi \dot\varphi 
 +\cdots\right]
\eeq
where dots denote terms without $\partial_0$, $h_{0i}^T$, or $\psi$. Now we can transform to the canonical formulation. Designating the conjugate momenta to $A_i^T$ and $\varphi$, respectively by $\pi_i^T$ and $\pi_\varphi$, the action in the canonical form becomes
\beq
S_{FP}^{(2)}=\mpl^2m^2\int d^4x \left[\frac{1}{2}{\dot A_i^T}\pi_i^T
 +\dot\varphi \pi_\varphi+\frac{1}{2}{h_{0i}^T}\pi_i^T +\cdots \right]\,.
\eeq
As expected $h_{0i}$ appears linearly and remains a Lagrange multiplier. On the other hand, $\varphi$, which is the longitudinal component of $A_i$, has now become dynamical through mixing with gravity.

Thus, the Fierz-Pauli term is a special combination that eliminates one of the six naively expected degrees of freedom at the linearized level. As Boulware and Deser showed, this in not generically true at nonlinear level--the sixth degree of freedom reappears and leads to a ghost instability. Therefore, although the fluctuations are healthy around Minkowski space (or more precisely when $g_{\mu\nu}=\eta_{\mu\nu}$ and $\phi^\mu=x^\mu$), they are not so around other macroscopically different backgrounds. This is of course not tolerable in a theory of gravity.

One may wonder if there exist constructions similar to that of Fierz and Pauli but at higher orders such that the sixth degree of freedom is completely eliminated. And indeed a two-parameter family of candidates (dRGT) was proposed by de Rham, Gabadadze, and Tolley based on an analysis in the decoupling limit, using a new St\"{u}eckelberg field.

\subsection{\label{newstuck}A New St\"{u}eckelberg Field}

We cannot do justice in explaining the concept of decoupling limit and its fruitfulness; the interested reader is referred to thorough discussions in \cite{AGS,giga1}. Here, and in the next subsection, we instead point out some relevant aspects and consequences of the decoupling limit analysis that led to dRGT.

Motivated by the fact that the longitudinal degree of freedom of a massive non-Abelian gauge field is the most strongly coupled component, Arkani-Hamed, Georgi, and Schwartz (AGS) introduced a new St\"{u}eckelberg field by decomposing $A_\mu$ as
\beq
\label{U1}
A_\mu=V_\mu+\partial_\mu\pi\,,
\eeq
so that the theory is furnished with a $U(1)$ gauge symmetry
\beq
\label{U12}
V_\mu\to V_\mu + \partial_\mu \alpha,\qquad \pi \to \pi - \alpha.
\eeq
In addition, since each $A_\mu$ field comes with one derivative, each $\pi$ carries two and the action acquires a Galilean symmetry $\pi\to \pi +b_\mu x^\mu+c$.

This new St\"{u}eckelberg field, which is supposed to describe the longitudinal scalar mode of massive gravity at high energies, is not as well-defined as the original scalar fields $\phi^A$ which were introduced in Sec. \ref{sec:scalar}, because the $\mu$ index in $A^\mu$ is a field-space index not a space-time one. However, as we argued above, around Minkowski background the fluctuations of the four scalar fields $A^\mu$ indeed  behave as components of a vector field. As long as we perform simultaneous Lorentz transformation in the space-time and field space, the two indices are indistinguishable. So it is legitimate to introduce $\pi$ as in \eqref{U1} to describe the longitudinal component. On the other hand on a generic background the meaning of $\pi$ can become very obscure.

In spite of this drawback the decomposition \eqref{U1} has been enormously fruitful in the study of massive gravity and greatly facilitates introduction and identification of the special features of dRGT. Later in Sec. \ref{sec:lin} we will show when this new St\"{u}eckelberg field does describe the longitudinal mode of graviton.

Substituting \eqref{U1} in the quadratic Lagrangian, one observes from \eqref{FF} that $\pi$ drops out of $\mathcal{L}_A^{(2)}$ up to total derivative terms because of its accidental $U(1)$ gauge symmetry, but $\mathcal{L}_{hA}^{(2)}$ gives a kinetic mixing between $h_{\mu\nu}$ and $\pi$
\beq
\label{Lhpi2}
\mathcal{L}_{h\pi}^{(2)}=-h^{\mu\nu}(\partial_\mu\partial_\nu-\eta_{\mu\nu}\Box)\pi\,.
\eeq
After integration by parts this is proportional to $\pi R^{(1)}$, where $R^{(1)}$ is the linearized Ricci scalar, and can be diagonalized by a shift of $h_{\mu\nu}$
\beq
h_{\mu\nu}\to h_{\mu\nu}+m^2\pi \eta_{\mu\nu}\,.
\eeq
As a result we get a mass term for $\pi$, mass mixing between $\pi$ and $h_{\mu\nu}$, kinetic mixing between $V_\mu$ and $\pi$, and most importantly a pure kinetic term for $\pi$ which, after restoring mass scales, looks like
\beq
\label{Lpi}
\mathcal{L}_\pi^{(2)}=\frac{3}{2}\mpl^2m^4\pi\Box\pi\,,
\eeq
and is the dominant piece at high momenta $p\gg m$. In fact, after canonically normalizing fields using the Einstein-Hilbert action \eqref{EH}, Eq. \eqref{FF} (with the overall $\mpl^2m^2$ restored), and \eqref{Lpi}, we get 
\beq
h^c_{\mu\nu}=\mpl h_{\mu\nu},\qquad V_c^\mu=\mpl m V^\mu\,,\qquad \pi_c = \mpl m^2\pi\,,
\eeq
and the high-energy, $m/p \to 0$, limit of the quadratic theory around Minkowski becomes 
\beq
\mathcal{L}^{(2)}\simeq \frac{1}{2}h_c\mathcal{E}h_c-\frac{1}{4}F_{\mu\nu}^2(V_c)+\frac{3}{2}\pi_c\Box\pi_c\,.
\eeq
The first term is the quadratic Einstein-Hilbert action describing two transverse-traceless tensor modes; the second is the square of the field strength of a $U(1)$ gauge field $V_c^\mu$, describing two transverse helicity-1 vectors, and the last term describes the scalar helicity-0 mode $\pi_c$, which is gauge invariant in this limit.

As anticipated, the self-interactions of $\pi$ become the most strongly coupled interactions; that is, after canonically normalizing the fields, these self-interactions are suppressed by lowest mass scales (cutoffs). For instance, the action \eqref{FPH} gives rise to interactions of the form
\beq
\label{L5}
\frac{(\partial \partial \pi_c)^3}{\Lambda_5^5},\qquad 
\frac{(\partial \partial \pi_c)^4}{\Lambda_4^8}\,,
\eeq
where $\Lambda_n\equiv(\mpl m^{n-1})^{1/n}$. The decoupling limit is a limit in which one keeps only the dominant interaction(s).

It was suggested by AGS and is very natural to remove these pure $\pi$ self-interactions by adding higher order terms to \eqref{FPH}. For instance, the gauge invariant form of $\mathcal{L}_A^{(2)}$ in \eqref{FF} and, consequently, the Fierz-Pauli structure are uniquely determined once we impose the requirement that the pure $\pi$ terms in the quadratic Lagrangian cancel up to a total derivative. Now, if there are higher order pure $\pi$ interactions there is always one or a few of them which dominate all nonlinearities. Around a macroscopic background they yield quadratic terms in $\pi$ which suggest the absence of the accidental gauge symmetry when $\mathcal{L}_A$ is expanded around that background \cite{Nicolis,Deffayet}. 

\subsection{dRGT}

Quite recently, de Rham, Gabadadze, and Tolley found this nonlinear completion of the Fierz-Pauli mass term \eqref{FPH}. The theory is an infinite series in terms of $H_{\mu\nu}$, since as a consequence of general covariance $H_{\mu\nu}$ is nonlinearly related to $\pi$. For instance, to cancel cubic terms in \eqref{L5} one introduces ${\cal O}(H_{\mu\nu}^3)$ terms which then give rise to sixth order terms in $\pi$, and so on. As becomes clear soon, the dRGT theory has only two free parameters, one at cubic and the other at quartic order.

dRGT contains couplings between $\pi$, $V_\mu$, and $h_{\mu\nu}$, such as
\beq
h(\partial \partial \pi)^2,\qquad \partial V \partial V \partial \partial \pi,\cdots
\eeq
which, after canonical normalization around Minkowski, are suppressed at least by powers of $\Lambda_3=(\mpl m^2)^{1/3}$ (hence the terminology $\Lambda_3$ theories). However, all pure $\pi$ self-interactions add up to total derivatives. The total derivative self-interactions of $\pi$ with two derivatives per field are closely related to Galilean interactions, and it was shown in \cite{Nicolis2} that in $d$ dimension there is a unique such term at each order up to $d$th order. Beyond $d$th order these total derivatives identically vanish. 

This means that, barring the tadpole (the cosmological constant), after fixing the coefficient of the quadratic term which determines the mass of graviton, one is free to choose $d-2$ coefficients for the $d-2$ higher order total derivatives. Beyond that, $\pi$ self-interactions must exactly cancel (without integration by parts) and there is no freedom. Since each $n$th order invariant monomial made out of $H^\mu_\nu$ has a unique structure of pure $\pi$ terms, which spans from $n$th to $2n$th order in $\pi$, after fixing the coefficient of the total derivatives the theory is fully specified. Hence, the two parameter family of dRGT in $4d$.

The fact that self-interactions of $\pi$ beyond quartic order must identically vanish, immediately implies that if in dRGT we replace all $A_\mu$ fields in $\mathcal{L}_A$ with $\partial_\mu\pi$, we must get zero beyond the quartic order. Therefore at those orders terms can be rearranged so that they contain at least one $F_{\mu\nu}=\partial_\mu A_\nu-\partial_\nu A_\mu$. 

However, writing $\partial_\mu A_\nu$ in terms of the symmetric and antisymmetric part
\beq
\partial_\mu A_\nu = (F_{\mu\nu}+S_{\mu\nu})/2\,,
\eeq
one realizes that the antisymmetry of $F_{\mu\nu}$ forces us to have either zero or at least two $F_{\mu\nu}$'s in each term. So each term beyond quartic order must contain at least two $F_{\mu\nu}$'s.

What about the lower order terms in $\mathcal{L}_A$? Well, we have already seen the example of the quadratic Lagrangian in \eqref{FF}--up to total derivatives it also contains two $F_{\mu\nu}$'s. The cubic Lagrangian is still easy to check explicitly. The only cubic self-interaction of $\pi$ that is a total derivative is 
\beq
\label{pi3}
(\Box \pi)^3-3\Box\pi(\partial_\mu\partial_\nu\pi)^2+2(\partial_\mu\partial_\nu\pi)^3\,,
\eeq
which means that apart from terms that contain two $F_{\mu\nu}$'s, the rest of the $\mathcal{L}_A^{(3)}$ should combine into a multiple of
\beq
\label{S3}
S_{\mu\mu}^3-3S_{\sigma\sigma}S_{\mu\nu}^2+2S_{\mu\nu}^3\,,
\eeq
to yield \eqref{pi3} upon substitution $A_\mu=\partial_\mu\pi$. However \eqref{S3} can be integrated by parts to give
\beq
12 \partial_\mu A_\nu F_{\mu\sigma}F_{\nu\sigma}-6\partial_\sigma A_\sigma F_{\mu\nu}^2+\text{total derivatives.}
\eeq

In Appendix \ref{app:fullder} we systematically prove that all combinations of the form \eqref{S3}, which yield a total derivative upon $A_\mu=\partial_\mu\pi$ replacement, reduce to terms with at least two $F_{\mu\nu}$'s after integration by parts. So fortunately we do not need to check $\mathcal{L}_A^{(4)}$ explicitly.

To summarize, the pure scalar sector $\mathcal{L}_A$ in dRGT can be written schematically as
\beq
\label{LA}
\mathcal{L}_A &=& -\frac{1}{4} FF[1+\partial A+\cdots]+\text{total derivatives.}
\eeq
Therefore, dRGT can be uniquely defined as a subclass of FP in which $\mathcal{L}_A$ contains at least two $F_{\mu\nu}$'s in each term, up to total derivatives.

Another property of dRGT which is again a direct consequence of removing pure $\pi$ interactions is the special form of the coupling between $h_{\mu\nu}$ and $\pi$ \cite{giga1,giga4}
\beq
\label{Lhpi}
\mathcal{L}_{h\pi}= h_{\mu\nu}\sum_{n=1}^{3} a_nX_n^{\mu\nu}(\Pi)+\mathcal{O}(h_{\mu\nu}^2)\,,
\eeq
where $\Pi_{\mu\nu}=\partial_\mu\partial_\nu\pi$ and the three $X_n^{\mu\nu}(\Pi)$ are given by
\beq
X_1^{\mu\nu}(\Pi)&=&{\varepsilon^{\mu}}^{\alpha\rho\sigma}
{{\varepsilon^\nu}^{\beta}}_{\rho\sigma}\Pi_{\alpha\beta}, \quad  \nonumber \\
X_2^{\mu\nu}(\Pi)&=&{\varepsilon^{\mu}}^{\alpha\rho\gamma}
{{\varepsilon^\nu}^{\beta\sigma}}_{\gamma}\Pi_{\alpha\beta}
\Pi_{\rho\sigma}, \nonumber \\
X_3^{\mu\nu}(\Pi)&=&{\varepsilon^{\mu}}^{\alpha\rho\gamma}
{{\varepsilon^\nu}^{\beta\sigma\delta}}\Pi_{\alpha\beta}
\Pi_{\rho\sigma}\Pi_{\gamma\delta}\,,
\label{Xs}
\eeq
[see Appendix \ref{app:hpi} for a derivation of \eqref{Lhpi} and \eqref{Xs}]. Note that $a_1h_{\mu\nu}X_1^{\mu\nu}(\Pi)$ is dictated by FP and must be equal to \eqref{Lhpi2}, so $a_1 = -1/2$. The other two coefficients $a_2,a_3$ are determined by the only two free parameters of the theory--the coefficients of the total derivative terms. 

As before, this form of $\mathcal{L}_{h\pi}$ in dRGT is very constraining, now for the scalar-metric coupling sector $\mathcal{L}_{hA}$. In order for $\mathcal{L}_{hA}$ to yield \eqref{Lhpi} upon decomposition $A_\mu=V_\mu+\partial_\mu\pi$, it must contain \eqref{Lhpi} with the replacement $\Pi_{\mu\nu}\to S_{\mu\nu}/2$. Also, the fact that there is no $X_n^{\mu\nu}(\Pi)$ beyond $n=3$ means that terms in $\mathcal{L}_{hA}$ which contain one $h_{\mu\nu}$ and more than three $A_\mu$'s identically vanish after $A_\mu = \partial_\mu\pi$ substitution. Therefore these terms can be rearranged such that each of them contain at least one $F_{\mu\nu}$.

So $\mathcal{L}_{hA}$ in dRGT is dictated to be of the form 
\beq
\label{LhA}
\mathcal{L}_{hA}=h_{\mu\nu}\sum_{n=1}^{3} a_nX_n^{\mu\nu}(S_{\mu\nu}/2)+h F(\partial A +\cdots)+\mathcal{O}(h_{\mu\nu}^2)\,,
\eeq
where the second type terms are written schematically.\footnote{I am thankful to Lasha Berezhiani for pointing out an error in the original formula.} We emphasize that after fixing the two free coefficients of dRGT, $\mathcal{L}_{A}$ and $\mathcal{L}_{hA}$ are completely determined but we do not need to know more than what is presented in \eqref{LA} and \eqref{LhA}.

We next study FP (and dRGT) around different backgrounds.

\section{\label{sec:lin}Linearized Background}

To understand the nature and properties of different degrees of freedom on a given background, one first needs to look at the quadratic action for fluctuations around it, as we did in Sec. \ref{sec:scalar} for Fierz-Pauli theory around Minkowski. To this end one can usually linearize the background since locally (for ultraviolet momenta) curvature effects are often negligible. We follow this route in this section but discover that, similar to the case of FP around Minkowski, the pure scalar sector of dRGT is degenerate around the linearized backgrounds. There is an accidental gauge symmetry which will be broken only through dynamical mixing with gravity. 

Consider a generic background of an FP theory \eqref{FPV}, given by $g^{u}_{\mu\nu}$ in the unitary gauge ($\phi^\mu=x_u^\mu$). Since the theory is generally covariant, to study the fluctuations around a given point $P$ we can perform a coordinate transformation to a local inertial frame of $P$. Taking $P$ as the origin of the new coordinate system the new background metric is given by
\beq
\label{hbg}
g^{bg}_{\mu\nu}=\eta_{\mu\nu}+h^{bg}_{\mu\nu}=\eta_{\mu\nu}-\frac{1}{6}(R_{\mu\alpha\nu\beta}+R_{\mu\beta\nu\alpha})x^\alpha x^\beta +\mathcal{O}(xxx)\,,
\eeq
and the background scalar fields $\phi_{bg}^\mu$ which are the unitary gauge coordinates can be Taylor expanded as
\beq
\label{phi}
\phi_{bg}^\mu=x_u^\mu(x)=x_u^\mu(0)+e^\mu_\nu x^\nu +\mathcal{O}(xx)\,,
\eeq
where the constant term $x_u^\mu(0)$ will be ignored in what follows since $\phi^\mu$ is derivatively coupled. In terms of $\phi_{bg}^\mu$ and the new metric \eqref{hbg} the unitary gauge inverse metric at point $P$ can be written as
\beq
\left. g_u^{\mu\nu}\right|_P=\left.g^{\alpha\beta}\partial_\alpha x_u^\mu\partial_\beta x_u^\nu
                                                        \right|_{x=0}
\left.=g^{\alpha\beta}\partial_\alpha\phi_{bg}^\mu\partial_\beta\phi_{bg}^\nu
                                       \right|_{x=0} 
   = \eta^{\alpha\beta}e^\mu_\alpha e^\nu_\beta\,.
\eeq
Thus the matrix $e^\mu_\nu$ appearing in \eqref{phi} is just the vielbein at point $P$. 

We now switch to the notation $A^\mu$ so that we can use the results of the last section and denote the fluctuations around the background \eqref{phi} by $a^\mu$
\beq
\label{Aa}
\phi^\mu= x^\mu-A^\mu=x^\mu-A^\mu_{bg}-a^\mu\,,
\eeq
where
\beq
\label{Abg}
A_{bg}^\mu=(\delta^\mu_\nu-e^\mu_\nu)x^\nu + \mathcal{O}(xx)\,.
\eeq

In this section we ignore $h^{bg}_{\mu\nu}$ [which is $\mathcal{O}(xx)$] and nonlinear corrections to $A_{bg}$ to obtain the curvature-independent contributions to the quadratic Lagrangian $\mathcal{L}^{(2)}$ for fluctuations of scalar fields $a^\mu$ and metric $h_{\mu\nu}$. Taking $h^{bg}_{\mu\nu}=0$ has two consequences: Firstly, $\mathcal{L}_a^{(2)}$, the part of $\mathcal{L}^{(2)}$ which is quadratic in $a^\mu$, receives contributions solely from $\mathcal{L}_A$ [see \eqref{LFP} for definition]. Secondly, $\mathcal{L}_{ha}^{(2)}$ is sourced by terms in $\mathcal{L}_{hA}$ which are linear in $h_{\mu\nu}$.

We first concentrate on $\mathcal{L}_a^{(2)}$ and distinguish two cases as follows.

\subsection{Symmetric Background}

When $\partial_\mu A^{bg}_\nu$ is symmetric (or equivalently $e_{\mu\nu}\equiv\eta_{\mu\sigma}e^\sigma_\nu=e_{\nu\mu}$ ) dRGT behaves in an interesting way because of its special form of $\mathcal{L}_A$ \eqref{LA}. The two common factors of $F_{\mu\nu}$ vanish when evaluated on the symmetric background, so they must be evaluated on perturbations and we get
\beq
\label{La2}
\mathcal{L}_a^{(2)}=B^{\mu\nu\alpha\beta}f_{\mu\nu}f_{\alpha\beta}\,.
\eeq
Here $B^{\mu\nu\alpha\beta}$ is a constant matrix depending on the background information which is encoded in $e^\mu_\nu$ and
\beq
f_{\mu\nu}\equiv\partial_\mu a_\nu-\partial_\nu a_\mu\,
\eeq
is the gauge invariant field strength of $a_\mu$. Consequently, the quadratic Lagrangian for $a_\mu$ has an accidental $U(1)$ gauge symmetry, and $a_0$ is nondynamical (there is no $\partial_0 a_0$). Therefore, as in Maxwell theory, $\mathcal{L}_a^{(2)}$ describes propagation of two transverse vector modes. 

In any other FP theory, $\mathcal{L}_A$ does not contain two $F_{\mu\nu}$'s per term and the quadratic Lagrangian for fluctuations would generically be of the form
\beq
\mathcal{L}_a^{(2)}\sim \partial a \partial a\,,
\eeq
without any gauge symmetry. These types of theories are easily seen to contain four degrees of freedom, one of them being a ghost (see, e.g., \cite{Lasha}).

\subsection{Nonsymmetric Background}

When $\partial_\mu A^{bg}_\nu$ is not symmetric ($e_{\mu\nu}\neq e_{\nu\mu}$), one can still use the symmetry of FP theories (\ref{FPV},\ref{Vphi}) under internal Lorentz transformation of $\phi^\alpha$ fields to symmetrize the background. The action depends on $\phi^\alpha$ only through the combination $g^{\mu\sigma}\partial_\sigma \phi^\alpha \partial_\nu \phi^\beta \eta_{\alpha\beta}$ and for any global Lorentz transformation $\Lambda^\alpha_\lambda$ we have
\beq
g^{\mu\sigma}\partial_\sigma (\Lambda^\alpha_\lambda\phi^\lambda) \partial_\nu (\Lambda^\beta_\gamma\phi^\gamma) \eta_{\alpha\beta}=g^{\mu\sigma}\partial_\sigma \phi^\lambda \partial_\nu \phi^\gamma \eta_{\lambda\gamma}\,.
\eeq

On the other hand an arbitrary matrix $e^\mu_\nu$ can be made symmetric (in the sense that $\eta_{\mu\sigma}e^\sigma_\nu=\eta_{\nu\sigma}e^\sigma_\mu$) by a Lorentz transformation on one of the indices
\beq
e^\mu_\nu\to \Lambda^\mu_\sigma e^\sigma_\nu\,,
\eeq
as long as it is not too far from $\delta^\mu_\nu$ (we will be more explicit about this in Sec. \ref{sec:ex}). Choosing $\Lambda^\mu_\nu$ to be this Lorentz transformation, we now have a set of transformed scalar fields $\tilde\phi^\mu=\Lambda^\mu_\nu \phi^\nu$. Defining, as before,
\beq
\tA^\mu=x^\mu-\tilde\phi^\mu\,,
\eeq
the background will be
\beq
\tA_{bg}^\mu=(\delta^\mu_\nu-\Lambda^\mu_\sigma e^\sigma_\nu)x^\nu\,,
\eeq
for which $\partial_\mu\tA^{bg}_\nu$ is symmetric and we are back to the previous case. Therefore, the quadratic Lagrangian $\mathcal{L}^{(2)}_\ta$ for the redefined field
\beq
\label{tildea1}
\ta^\mu=\Lambda^\mu_\nu a^\nu\,
\eeq
will have a gauge symmetry in dRGT and describes two transverse vector modes, but not in any other FP. Before turning to the analysis of mixing with gravity, it is well-timed to make a few comments:

\begin{itemize}

\item The symmetrization of the background can be achieved, equally well, by using the freedom in choosing a Lorentz transformed local inertial frame. To symmetrize $\partial_\mu\phi^\nu$ one can either do a rotation (or boost) in field space or an opposite one in coordinate space. 

\item We mentioned earlier that switching from Latin to Greek indices, {\it per se}, does not change the nature of the fields. The important steps that packaged fluctuations of the four scalar fields $a^A$ (back to Latin indices!) into a vector field were (a) we transformed to the local inertial frame in which the space-time metric becomes the same as the internal metric $\eta_{AB}$. As a result, the field indices and the space-time indices see the same metric, as was the case for FP around Minkowski. 

However, there is still a big difference: now we have another nontrivial matrix $\partial_\mu\phi_{bg}^A$, or better said $\partial_\mu\phi_A=\eta_{AB}\partial_\mu\phi^B$. (b) By rotating field space basis relative to coordinate basis we symmetrized $\partial_\mu\phi_A$ so that it does not differentiate among space-time and field indices anymore. Only then can we interpret Greek indices on $\tilde a^\mu$ as space-time indices and the field itself as a space-time vector. This property is preserved under simultaneous Lorentz transformations in field and coordinate space.

\item Now we can see when it is legitimate to do the decomposition 
\beq
a_\mu=v_\mu+\partial_\mu \varpi\,.
\eeq
This is possible after transformation to the locally inertial frame (or more generally matching the space-time metric and the reference metric in any other way) and symmetrization of the background, i.e. when $a_\mu$ becomes a space-time vector. 

\item Now consider the original St\"{u}eckelberg field which described the longitudinal mode around Minkowski
\beq
\label{U11}
A_\mu=V_\mu+\partial_\mu \pi\,,
\eeq
and suppose we are given a background solution $h^{bg}_{\mu\nu}$, $V^{bg}_\mu$, and $\pi_{bg}$. The question we wish to answer is, when does the perturbation $\delta\pi$ around $\pi_{bg}$ describe the longitudinal mode of graviton around this background? First of all, we see that going to the local inertial frame is impossible since $a_\mu=\delta A_\mu$ behave as four scalar fields under general coordinate transformation while $\partial_\mu\delta\pi$ behaves as a vector field. Therefore, $h^{bg}_{\mu\nu}$ must be small such that different indices see approximately the same metric. Secondly, the background should be symmetric. An important example is when we have an almost purely $\pi$ background, such as the spherically symmetric asymptotically flat solution in the Schwarzschild frame and when $r\gg r_g$ \cite{David}. In these situations the background is trivially symmetric 
\beq
\partial_\mu\partial_\nu \pi_{bg}=\partial_\nu\partial_\mu \pi_{bg}\,,
\eeq
and $\delta\pi$ describes the longitudinal mode of the vector $a_\mu$ and the graviton.

\item Our analysis showed that unlike generic FP where $\mathcal{L}_a^{(2)}$ on a generic linearized background describes four degrees of freedom, in dRGT $\mathcal{L}_a^{(2)}$ describes two. It has an accidental $U(1)$ gauge symmetry and $a_0$ is nondynamical. Therefore, as in the case of FP around Minkowski, the helicity-0 degree of freedom should get its kinetic term via dynamical mixing with gravity $\mathcal{L}_{ha}^{(2)}$. 

\end{itemize}

\subsection{\label{mixing}Mixing with Gravity}

Since we are in the local inertial frame and the curvature effects have been temporarily ignored, the quadratic mixing between $a_\mu$ and $h_{\mu\nu}$, comes entirely from terms in $\mathcal{L}_{hA}$ that are linear in $h_{\mu\nu}$. This makes the analysis very simple because as shown in \eqref{LhA} these terms are constrained in dRGT to be of the form
\beq
\label{LhA3}
\mathcal{L}_{hA}=h_{\mu\nu}\sum_{n=1}^{3} a_nX_n^{\mu\nu}(S_{\rho\sigma}/2)+h F(\partial A +\cdots)+\mathcal{O}(h_{\mu\nu}^2)\,.
\eeq
The first term $h_{\mu\nu}X_1^{\mu\nu}(S_{\rho\sigma}/2)$ is what we studied in the case of FP around Minkowski \eqref{LhA2}. We saw that the symmetry of the indices of $X_1^{\mu\nu}$ is such that $\partial_0 A_0$ is multiplied by spatial components of the metric fluctuation $h_{ij}$. Therefore, we could integrate by parts the time derivative of $\partial_0 A_0$ to make $A_0$ a nondynamical field without making the Lagrange multipliers of GR, $h_{00}$ and $h_{0i}$, dynamical. It follows that the longitudinal component of $A_i$ becomes dynamical through mixing and we get three new degrees of freedom in addition to the two tensor modes of the metric.

For a nontrivial symmetric background, the analysis of the first term does not change at all; we just need to replace $A_\mu$ with $a_\mu$. But, another contribution to $\mathcal{L}_{ha}^{(2)}$ comes from equating, respectively, one and two $S_{\mu\nu}$'s in $X_2(S/2)$ and $X_3(S/2)$ to their background values $S^{bg}_{\mu\nu}=2\partial_\mu A^{bg}_\nu$. It again follows from the antisymmetry of the Levi-Civita symbols in the definition of $X_n^{\mu\nu}$ that the terms with $\partial_0 a_0$ cannot contain $h_{00}$ or $h_{0i}$. More explicitly, we have from these terms
\beq
\mathcal{L}_{hX}^{(2)}\sim h_{\mu\nu}\vep^{\mu\alpha\cdots}\vep^{\nu\beta\cdots}\partial_\alpha a_\beta\cdots\,,
\eeq
where we have suppressed irrelevant indices and background fields. So, $a_0$ remains nondynamical.

The only remaining part of $\mathcal{L}_{ha}^{(2)}$ comes from the second type terms in \eqref{LhA3}, where the factor of $F_{\mu\nu}$ vanishes on the symmetric background and must be evaluated on perturbations. The contribution of these terms is, therefore, of the form $h f$ in which $a_0$ is manifestly nondynamical without any need for integration by parts.

Note finally that the presented proof of ghost freedom around linearized backgrounds goes beyond what is already known in the decoupling limit of dRGT. In common between the two analyses is that they are applicable only if curvature is negligible so that $h_{\mu\nu}$ can be kept linearly. However, the decoupling limit truncation of dRGT is proven to be ghost free only around pure $\pi$ backgrounds, while in the presence of generic background vector fields it is not a consistent truncation of the theory and propagates a sixth degree of freedom.

\subsection{\label{sec:ex} An Example}

To illustrate the power and limitations of our approach, let us study a simple but nontrivial example, where the invariant interval in the unitary gauge is given by
\beq
\label{bg}
ds^2=g^u_{\mu\nu}dx_u^\mu dx_u^\nu=-dt_u^2+\delta_{ij}(dx_u^i+2l^i dt_u)(dx_u^j+2l^j dt_u)\,.
\eeq
This background was first considered in \cite{Slava}, and it was claimed that the fluctuations of all four scalar fields become dynamical around this background. Later it was shown in \cite{giga_stu} that via a field redefinition, one of the fields can be made nondynamical and it was argued that the pure scalar sector propagates three degrees of freedom based on an analysis of Hamiltonian constraints.

We now analyze this problem by going to the local inertial frame as prescribed in previous subsections. Note that the unitary gauge metric \eqref{bg} is flat and can be transformed to Minkowski metric in a new coordinate system $(t,x^i)$ if we take
\beq
t_u&=&\phi_{bg}^0=t\,,\\
x^i_u&=&\phi_{bg}^i=x^i-2l^it\,.
\eeq
This is perfectly linear, but not symmetric: $\partial_i\phi_{bg}^0=0$ while $\partial_0\phi_{bg}^i=-2l^i$. To symmetrize it, we can perform a Lorentz boost with parameter $l^i$ 
\beq
\label{boost}
\tilde\phi^0=\frac{1}{\sqrt{1-l_k^2}}(\phi^0+l_i\phi^i)\,,\qquad
\tilde\phi^i=\frac{1}{\sqrt{1-l_k^2}}(\phi^i+l^i\phi^0)\,.\qquad
\eeq
It can be easily checked that this transformation makes $\partial_\mu\tilde\phi^{bg}_\nu=\eta_{\nu\sigma}\partial_\mu\tilde\phi^\sigma_{bg}$ symmetric. Therefore, the fluctuations $\tilde a^\mu$ form a vector field. As shown before, the pure scalar sector $\mathcal{L}_{\tilde a}^{(2)}$ in dRGT will depend on $\tilde a_\mu$ only through its field strength $\tilde f_{\mu\nu}$, therefore it describes propagation of two transverse vectors. The third mode becomes dynamical solely through mixing with gravity.

Now we can see what it means to be too far from Minkowski to symmetrize the background. It amounts to taking $l_k^2>1$ in \eqref{boost}. That is, unlike Euclidean space where any vector can be rotated to align with any other, in a space with Lorentzian signature, boosts have some limitations and cannot align a timelike vector with a spacelike one. The unitary-gauge metrics with a nonzero shift vector need a boost for symmetrization, so those with a very large shift vector are expected to be nonsymmetrizable.

Let us see if the Hamiltonian analysis of Hassan and Rosen \cite{Rachel} can resolve this issue. In that approach one wants to redefine the shift vector $N^i$ in terms of a new vector $n^i$, such that the action becomes linear in the lapse $N$. Since the GR action contains $N^i$ linearly, the field redefinition should be linear in $N$. Consider for simplicity a $2d$ space-time; then the redefinition can be written as
\beq
\label{redef}
N^1=(1+ND)n^1\,.
\eeq
It can be shown using a representation of dRGT \cite{Rachel_resum} that the coefficient $D$ must satisfy
\beq
\label{D}
\sqrt{1-(n^1)^2}D=\sqrt{\gamma^{-1}-(Dn^1)^2}\,,
\eeq
where $\gamma$ is the $1d$ spatial metric. The solution is $D=\sqrt{\gamma^{-1}}$ and using the two dimensional analogue of \eqref{bg} which has $\gamma=1$ and $N=1$, we get $D=1$. Substitution in (\ref{redef}) gives $n^1=N^1/2$. Thus, quite similarly to the symmetrization \eqref{boost}, when $N^1$ is so large that $(n^1)^2\geq 1$, the redefinition \eqref{redef} is ill-defined since the prefactor $\sqrt{1-(n^1)^2}$ in \eqref{D} becomes complex. It appears that these metrics, if realized as background solutions, evade any proof of ghost freedom. 

The resolution, however, is provided by Hassan and Rosen. The equations of motion for $n^i$ ensure that, as long as the unitary gauge is well-defined, $(n^i)^2$ does not exceed $1$ and the field redefinition is possible. Therefore, the nonsymmetrizable ``backgrounds'' which are characterized by unitary-gauge metrics with very large values of shift vector are not true background solutions of the theory.\footnote{I am thankful to Rachel Rosen for explaining this to me.}

We will not further investigate whether all nonsymmetrizable backgrounds fall into this category. In what follows we will focus on the continuum of backgrounds which are close enough to Minkowski to be symmetrized.

\section{\label{sec:nl}Nonlinear Background}

In this section we generalize the analysis of the last section to nonlinear (or curved) backgrounds. The general procedure should have become clear by the intuition we gained from linear backgrounds. The field space basis must be rotated point by point so that the field and coordinate indices are treated symmetrically. Then the fluctuations $a^\mu$ form a vector field. 

We break the proof into two steps. First we ignore curvature but allow for nonlinearities in $\phi^\mu_{bg}$. This can be thought of as a locally flat space but with nonzero gravitational acceleration. We next include curvature.

\subsection{\label{sec:zerocurv}Zero Curvature}

Let us try to follow the steps of the last section and see when modifications are needed. As before we can coordinate transform to a local inertial frame of an arbitrary point and set $h_{\mu\nu}^{bg}=0$ in its vicinity. The background scalar fields are, in general, nonlinear so that $\partial_\mu \phi_{bg}^\alpha$ is space-time dependent. If this matrix turns out to be symmetric in a region much larger than the de Broglie wavelength of the modes in question then as before we conclude that the quadratic Lagrangian for perturbations $\mathcal{L}_a^{(2)}$ acquires an accidental gauge symmetry and describes only two vector degrees of freedom.

If the background is not symmetric, it means that it differentiates among coordinate and field indices. We then perform an internal Lorentz transformation of the field space basis to align them with the coordinate basis, thereby symmetrizing the background. Generically, however, a constant Lorentz transformation is unable to do the job and we need to use the larger symmetry of the theory under local Lorentz transformations $\Lambda(x)$
\beq
g^{-1}(\partial \phi) \eta (\partial \phi)^T=g^{-1}(\partial \phi)\Lambda(x)\eta \Lambda^T(x)(\partial \phi)^T\,.
\eeq
Barring backgrounds that are too far from Minkowski we can find an appropriate $\Lambda(x)$ to align the two bases and symmetrize the background in the whole region so that
\beq
[(\partial \phi_{bg})\Lambda(x)\eta]^T=(\partial \phi_{bg})\Lambda(x)\eta \,.
\eeq
Having symmetrized the background, we can now return to the expansion of the theory in terms of $A^\mu$. The theory depends on $\phi^\mu$ only through first derivatives, $\partial_\nu\phi^\mu$; therefore, when we defined $A^\mu=x^\mu-\phi^\mu$ and expanded the action in terms of $A^\mu$, it could be thought of as an expansion in $\delta^\mu_\nu-\partial_\nu\phi^\mu$. Thus, after the above transformation, $\partial_\mu A^\nu$ must be replaced by
\beq 
\label{Ldphi}
\partial_\mu A^\nu\to \delta^\nu_\mu - \Lambda^\nu_\sigma(x)\partial_\mu\phi^\sigma\,.
\eeq
We will see that the $x$ dependence of $\Lambda^\nu_\sigma(x)$ leads to some modifications in the analysis of nonlinear backgrounds which make them fundamentally different from the linear ones.

\subsubsection{\label{sec:diag}Diagonalization of the Perturbations}

First consider the perturbations $a^\mu$ of the scalar fields. According to \eqref{Ldphi}, we should replace $\partial_\mu a^\nu$ with $\Lambda^\mu_\sigma(x)\partial_\nu a^\sigma$ in $\mathcal{L}_A$ and $\mathcal{L}_{hA}$. When $\Lambda^\mu_\sigma$ was a constant, it could be passed inside the derivative and the perturbations $a^\mu$ could be diagonalized by defining
\beq
\label{tildea}
\tilde a^\mu=\Lambda^\mu_\sigma a^\sigma\,,
\eeq
where $\tilde a^\mu$ is a true vector field. In the case of $x$-dependent $\Lambda^\mu_\nu(x)$, the same procedure yields an extra piece
\beq
\label{Lambda_da}
\Lambda^\mu_\sigma(x)\partial_\nu a^\sigma=\partial_\nu(\Lambda^\mu_\sigma(x) a^\sigma)-\left(\partial_\nu \Lambda^\mu_\sigma(x)\right)a^\sigma=\partial_\nu\tilde a^\mu+C^\mu_{\nu\sigma}\tilde a^\sigma\,,
\eeq
where in the last step we have defined
\beq
\tilde a^\mu=\Lambda^\mu_\nu(x)a^\nu\,,\qquad\text{and}\qquad
C^\mu_{\nu\sigma}=-\left(\partial_\nu \Lambda^\mu_\rho(x)\right)\Lambda^\rho_\sigma\,.
\eeq
We emphasize that the above space-time dependent field redefinition is the manifestation of the point by point alignment of coordinate and field bases. 

The presence of the extra piece in \eqref{Lambda_da} makes an important difference: Suppose we want to derive $\mathcal{L}_a^{(2)}$, the quadratic Lagrangian for the fluctuations in the pure scalar sector. In the last section we saw that after symmetrization and field redefinition \eqref{tildea}, $\mathcal{L}_{\tilde a}^{(2)}$ in dRGT only contains the antisymmetric field strength $\tilde f_{\mu\nu}=\partial_{[\mu} \tilde a_{\nu]}$. That statement really meant that, upon lowering by Minkowski metric, the indices $\mu$ and $\nu$ of $\Lambda^\mu_\sigma(x)\partial_\nu a^\sigma$ are antisymmetrized. This gives $\tilde f_{\mu\nu}$ for constant $\Lambda^\mu_\sigma$; however, in general there is an additional term
\beq
\label{[Lda]}
\Lambda_{\mu\sigma} \partial_{\nu}a^\sigma-\Lambda_{\nu\sigma} \partial_{\mu}a^\sigma=\tilde f_{\mu\nu}+C_{[\mu\nu]\sigma}\tilde a^\sigma\,.
\eeq
Therefore unlike linear backgrounds where $\mathcal{L}_{\tilde a}^{(2)}$ was formed out of two $\tilde f_{\mu\nu}$'s and had an accidental gauge symmetry, here we schematically have
\beq
\label{Lta}
\mathcal{L}_{\ta}^{(2)}\sim \tf\tf+C \tf\ta+CC\ta\ta\,,
\eeq
where the indices are suppressed. This Lagrangian looks a lot like that of a massive gauge field in which $\ta_0$ is nondynamical, since there is no $\partial_0\ta_0$, and three degrees of freedom propagate; two transverse helicity-1, and a longitudinal helicity-0. So, the nonlinearities of the background, which are characterized by $C^\mu_{\nu\sigma}$, break the degeneracy of the pure scalar sector and give a contribution to the kinetic term of the helicity-0 mode which used to get its dynamics through mixing with gravity.

\subsubsection{\label{sec:ninteg}NonIntegrability of the Background}

Next we consider the symmetrized background. According to \eqref{Ldphi} we should replace $\partial_\nu A^\mu_{bg}$ with the following symmetric matrix
\beq
\label{(dA)}
(d\tA_{bg})_\nu^\mu\equiv \delta^\mu_\nu-\Lambda^\mu_\sigma(x)\partial_\nu\phi_{bg}^\sigma\,.
\eeq
Again, when $\Lambda^\mu_\sigma$ is constant, we can commute it with the derivative and define a new background field $\tA_{bg}^\mu=x^\mu-\Lambda^\mu_\sigma\phi_{bg}^\sigma$ whose derivative is \eqref{(dA)}. However, such a new background field does not necessarily exist when $\Lambda^\mu_\sigma$ depends on $x$; that is $(d\tA_{bg})$ is nonintegrable. 

One may wonder whether this causes any technical obstruction in the analysis. Let us recall that the technical consequence of symmetrization was to make $F_{\mu\nu}$'s in the expansions $\mathcal{L}_A$ and $\mathcal{L}_{hA}$ vanish on the background so that they had to be evaluated on perturbations. But the same will be true if we replace $\partial_\mu A_\nu$ inside $F_{\mu\nu}=\partial_{[\mu}A_{\nu]}$ with any symmetric matrix, such as our $(d\tA_{bg})_{\mu\nu}$; it still vanishes. So we still need to evaluate all $F_{\mu\nu}$'s on the perturbations, which in the present case are $\Lambda_{\mu\sigma}(x)\partial_\nu a^\sigma$ and the result was discussed in section \ref{sec:diag}.

The only technical difference caused by the nonintegrability of the background would arise when we integrated some derivatives by parts to get the desired form of $\mathcal{L}_A$ with two $F_{\mu\nu}$'s per term. This is not permitted anymore and one needs to look at the original form of $\mathcal{L}_A$. Fortunately, integration by parts was needed only for a finite number of terms, namely, the quadratic, cubic, and quartic terms that upon substitution $A_\mu=\partial_\mu\pi$ would yield total derivatives. We will systematically treat these terms in Appendix \ref{app:quad} and confirm that their contribution to $\mathcal{L}_a^{(2)}$ is still of the form \eqref{Lta}.

Finally, we should make sure that the structure of mixing terms in $\mathcal{L}_{h\ta}^{(2)}$ does not change significantly; to be more precise, we should show that no $\partial_0\ta_0$ multiplies $h_{00}$ and $h_{0i}$. Since $h^{bg}_{\mu\nu}=0$, we only need terms in $\mathcal{L}_{hA}$ which are linear in $h_{\mu\nu}$ [Eq. \eqref{LhA}],
\beq
\label{LhA4}
\mathcal{L}_{hA}=h_{\mu\nu}\sum_{n=1}^{3} a_nX_n^{\mu\nu}(S_{\rho\sigma}/2)+h F(\partial A +\cdots)+\mathcal{O}(h_{\mu\nu}^2)\,.
\eeq
Note that we never used integration by parts to obtain this; therefore, after symmetrization $F_{\mu\nu}$ in the second part must be replaced by $\tilde f_{\mu\nu}+C_{[\mu\nu]\sigma}\tilde a^\sigma$ because $(d\tA_{bg})_{[\mu\nu]}=0$. Hence there is no $\partial_0 \ta_0$ in this part. One then needs to take one of the $\partial_\mu A_\nu$'s in each $X_n$ and replace it by a perturbation, which now reads $\partial_\mu\ta_\nu+C_{\mu\nu\sigma}\ta^\sigma$. The $\partial_\mu\ta_\nu$ part was encountered before; when $\mu=\nu=0$ it will not multiply $h_{00}$ or $h_{0i}$ because of the symmetry properties of $X_n^{\mu\nu}$. The $C\ta$ term does not carry any derivatives.

\subsection{Curved Background}

Only one nontrivial step is required to generalize the proof to include metric curvature. It should have become evident by now that the key ingredient of the proof is alignment of field space basis and coordinate basis. So far, an easy step towards this alignment was going to the locally inertial frame where the background space-time metric and the field space metric $\eta_{AB}$ become the same (in this section it will be equally convenient to use Latin indices for scalar fields). In a curved space-time, this can be done only at a single point and in its vicinity the two metrics start to deviate from each other according to
\beq
\label{gbg}
g^{bg}_{\mu\nu}=\eta_{\mu\nu}+h^{bg}_{\mu\nu}=\eta_{\mu\nu}-\frac{1}{6}(R_{\mu\alpha\nu\beta}+R_{\mu\beta\nu\alpha})x^\alpha x^\beta +\mathcal{O}(xxx)\,.
\eeq

This issue can be overcome by a technically simple but conceptually fundamental twist in our approach. Instead of attempting to align the curved space-time metric with the flat field space metric, we do the opposite by introducing vielbeins in the field space. Let us call the background metric $\bar g_{\mu\nu}$
\beq
\bar g_{\mu\nu}(x)= g^{bg}_{\mu\nu}(x)\,,
\eeq
we can find a set of vielbeins that relate the original flat reference metric $\eta_{AB}$ to $\bg_{\mu\nu}$ as follows
\beq
\label{eta}
\eta_{AB}=e^\sigma_A(x)e^\rho_B(x)\bg_{\sigma\rho}(x)\,.
\eeq
As usual, one has freedom in the choice of the vielbein and the different choices are related by
\beq
\label{eGamma}
e^\mu_A\to \Gamma^\mu_\nu e^\nu_A\,,
\eeq
where $\Gamma^\mu_\nu$ are generalized Lorentz transformations which preserve the metric $\bg_{\mu\nu}$:
\beq
\label{Gamma}
\Gamma^\sigma_\mu\Gamma^\rho_\nu \bg_{\sigma\rho}=\bg_{\mu\nu}\,.
\eeq
They form a group which is isomorphic to the Lorentz group $SO(1,3)$. 

Once the substitution \eqref{eta} is done in the FP action
\beq
S_{FP}=-\frac{1}{4}\mpl^2m^2\int d^4x\sqrt{-g}V\left(g^{\mu\sigma}e^\rho_A\partial_\sigma\phi^A e^\lambda_B\partial_\nu\phi^B\bg_{\rho\lambda}\right)\,,
\eeq
it can still be expanded as a sum of three pieces
\beq
\label{Lbar}
S_{FP}=\mpl^2m^2\int d^4x\sqrt{-\bg}(\mathcal{\bar L}_h+\mathcal{\bar L}_{h\bA}+\mathcal{\bar L}_\bA)\,,
\eeq
where $\mathcal{\bar L}_h,\mathcal{\bar L}_{h\bA},$ and $\mathcal{\bar L}_\bA$ are, respectively, the Lagrangians for the metric, coupling between metric and the scalar fields, and pure scalar fields. They are the same as $\mathcal{L}_h$, $\mathcal{L}_{hA}$, and $\mathcal{L}_A$, except for three modifications: (a) Instead of Minkowski metric, all indices are contracted by $\bg_{\mu\nu}$ and its inverse. (b) The metric perturbation $h_{\mu\nu}$ is with respect to the new reference metric 
\beq
g_{\mu\nu}=\bg_{\mu\nu}+h_{\mu\nu}\,.
\eeq
And (c) all $\partial_\mu A_\nu$'s should be replaced with
\beq
\label{dA}
\partial_\mu A_\nu\to \bg_{\mu\nu}-\bg_{\nu\sigma}e^\sigma_A\partial_\mu \phi^A\,.
\eeq
The background value of \eqref{dA} can be symmetrized by exploiting the freedom \eqref{eGamma} in the choice of $e^\sigma_A$. The result is a symmetric, but not necessarily integrable, matrix
\beq
(d\bA_{bg})_{\mu\nu}=\bg_{\mu\nu}-\bg_{\nu\sigma}e^\sigma_A\partial_\mu \phi_{bg}^A\,.
\eeq
We can now diagonalize the perturbations of the scalar fields $a^A$ by defining the locally aligned perturbations $\ba_\mu$ and the connection matrix $C_{\mu\nu}^\sigma$ such that
\beq
\label{geda}
\bg_{\nu\sigma}e^\sigma_A\partial_\mu a^A=\partial_\mu\ba_\nu+C_{\mu\nu}^\sigma\ba_\sigma\,,
\eeq
where
\beq
\ba_\mu=\bg_{\mu\sigma}e^\sigma_A a^A\,,\qquad\text{and,}\qquad C_{\mu\nu}^\sigma=-\partial_\mu(\bg_{\nu\lambda}e^\lambda_A(x))\bg^{\rho\sigma}{e^{-1}}^A_\rho\,.
\eeq

Now we are back to the situation of the previous subsection except for two minor caveats. In the analysis of the quadratic, cubic, and quartic terms in $\mathcal{L}_A$ which give $F_{\mu\nu}$ only after integration by parts, we should also include the nonflatness of the metric. However, this cannot qualitatively change our previous conclusion based on a flat metric, because non-flat metric does not change the symmetry of indices and its only effect is to introduce an $x$-dependent coefficient. But $x$-dependent coefficients were already present before (see Appendix \ref{app:quad}). Therefore, $\mathcal{L}_\ba^{(2)}$ will again be of the form \eqref{Lta}.

The other caveat is that for a generic $\bg_{\mu\nu}$, the position of indices matters. The dynamical variables of GR are $g_{ij}$ and at the linearized level $h_{ij}$. So we should make sure that in $\mathcal{L}_{ha}^{(2)}$, $\partial_0\ba_0$ is multiplied by $h_{ij}$ with lower indices. This is evident from the explicit form of $\mathcal{L}_{hA}$, where $\partial_0 \ba_0$ appears only in the combination
\beq
\mathcal{L}_{hX}^{(2)} \sim h_{\mu\nu}\vep^{\mu\alpha\cdots}\vep^{\nu\beta\cdots}\partial_\alpha \ba_\beta\cdots \,,
\eeq
and we have suppressed irrelevant indices and background matrices. This completes our proof of ghost freedom in dRGT massive gravity.

We conclude this section with two remarks. First, that in our final approach the special choice of Minkowski reference metric, $\eta_{AB}$, did not play any role. For an arbitrary space-time dependent reference metric $\gamma_{AB}(x)$ one could define generalized vielbeins
\beq
\gamma_{AB}(x)=e^\sigma_A(x)e^\rho_B(x)\bg_{\sigma\rho}\,,
\eeq
and the rest of the proof would proceed unaltered. Therefore, as first shown in \cite{Rachel_ref}, dRGT with any reference metric propagates five degrees of freedom.

Secondly, the action \eqref{Lbar} provides the right framework to consistently generalize the definition of decoupling limit to cases where the background metric is different from the reference metric of FP. The St\"{u}eckelberg field $\pi$ can again be defined to describe the longitudinal mode of the true vector field $\bar a_\mu$, and the hierarchy of interactions becomes evident once different helicity modes are canonically normalized.

\section{\label{sec:disp}Dispersion Relations and the Cutoff}

Having established the number of degrees of freedom in dRGT to be five, one can come up with a more practical method of deriving dispersion relations for helicity-1 and helicity-0 modes. We saw that around linearized backgrounds the pure scalar sector is nondegenerate for transverse vector (helicity-1) modes. It is also easy to see that mixing with gravity has a negligible effect on them at large momenta $p\gg m$. The relevant quadratic Lagrangian for these modes looks like
\beq
\label{Lv2}
\mathcal{L}^{(2)}\sim \mpl^2m^2(ff+h\partial a)\,,
\eeq
where $h_{\mu\nu}$ and $a^\mu$ are not canonically normalized. Normalizing them by taking $a^\mu_c=\mpl m a^\mu$ and $h^c_{\mu\nu}=\mpl h_{\mu\nu}$, and assuming that canonically normalized variables are of the same order of the energy scale of interest confirms the above statement.

Therefore, investigating the behavior of transverse vector modes is relatively easy. One can first linearize any background (since the nonlinearities introduce subleading corrections), then align the field and coordinate bases (for instance, by going to the local inertial frame and symmetrizing), and finally freeze the metric and just study the transverse excitations of the pure scalar theory such as $a_y(t,x)$ (perhaps using the resummed version of dRGT \cite{giga2}). 

Some of the above steps may often be unnecessary. For instance, whenever symmetrization of the background does not affect the mode in question, say when we study $a_y$ and the symmetrizing Lorentz transformation is a boost in the $t-x$ plane, we do not even need to symmetrize the background. Nevertheless, symmetrization usually simplifies the computation. See \cite{Andrei,giga_suplum} for an example. 

We should remark that the coefficients and the matrices that contract indices in \eqref{Lv2} depend on the background solution. Typically they are of the order of $\partial_\mu\phi_{bg}^\nu$ which is $\delta^\nu_\mu$ on Minkowski background but can deviate significantly for different solutions (for instance, through Vainshtein mechanism \cite{stars}). Examples in which the vector modes become ghostlike or infinitely strongly coupled can be found in \cite{giga4,KoyamaSA}.

\subsection{\label{scalar}The Scalar Mode}

Finding the dispersion relation of the scalar mode needs more effort. On linear backgrounds the pure scalar sector of dRGT is degenerate and nonlinearities are important. We saw that these nonlinearities generate mass for the properly aligned vector field $\ba_\mu$, which provides dynamics for the longitudinal mode. However, so far it seemed impractical to find the explicit form of this term. In this section we introduce a method to find the leading mass term. 

As already mentioned, since now $\ba_\mu$ is a vector field one can safely define the St\"{u}eckelberg field $\pi$
\beq
\label{dpi}
\ba_\mu=\bv_\mu+\partial_\mu \pi\,,
\eeq
which captures the dynamics of the longitudinal mode at high momenta. The contribution of mixing with gravity to the dispersion relation of $\pi$ was discussed in Secs. \ref{newstuck} and \ref{mixing}. The recipe to find the contribution of nonlinearities around a given point $P$ is

\begin{enumerate}

\item Coordinate transform to a local inertial frame where $P$ is located at $x^\mu=0$. The metric will look like
\beq
\label{hbg1}
g^{bg}_{\mu\nu}=\eta_{\mu\nu}+h^{bg}_{\mu\nu}=\eta_{\mu\nu}-\frac{1}{6}(R_{\mu\alpha\nu\beta}+R_{\mu\beta\nu\alpha})x^\alpha x^\beta +\mathcal{O}(xxx)\,.
\eeq

\item Taylor expand the scalar field background $\phi^\mu_{bg}$ and symmetrize the linear part to obtain
\beq
\label{phixx}
\phi^\mu_{bg}=e^\mu_\nu x^\nu -\frac{1}{2}C^\mu_{\alpha\beta}x^\alpha x^\beta+\mathcal{O}(xxx)\,,
\eeq
where after lowering the indices by Minkowski metric $e_{\mu\nu}$ is symmetric. We denote the fluctuations of $\phi^\mu$ by $-a^\mu$. Note that $a^\mu$ is different from a fully aligned vector field $\ba^\mu$ only by terms that disappear at $x=0$. We exploit this fact to obtain the leading quadratic Lagrangian $\mathcal{L}_\ba^{(2)}$ from the more accessible one $\mathcal{L}_a^{(2)}$. By leading we mean the part that survives $x\to 0$ limit.

\item Neglecting $h^{bg}_{\mu\nu}$ for the moment, one can expand the action to second order in $a^\mu$ around the background \eqref{phixx}. For this purpose the resummed version of the theory can be useful. It will become clear that one can keep only the linear and quadratic terms in \eqref{phixx}, so the result schematically looks like
\beq
\label{L2}
\mathcal{L}_{a}^{(2)}\sim \partial a \partial a+ C x\partial a \partial a+CCxx\partial a \partial a\,,
\eeq
plus terms which are higher order in $x$ and therefore subleading. Since $h_{\mu\nu}^{bg}$ was neglected, what we have obtained is merely the contribution of the pure scalar sector $\mathcal{L}_A$ to $\mathcal{L}_a^{(2)}$.

\item Since no $x$-dependent symmetrization has been performed, we can use the fact that $\mathcal{L}_A$ contains two $F_{\mu\nu}$'s per term up to total derivatives. $F_{\mu\nu}$ on the linear part of the background which is $\partial_\mu A^{lin.bg}_\nu=\eta_{\mu\nu}-e_{\mu\nu}$ vanishes since $e_{\mu\nu}$ is symmetrized. However, the nonlinear part is not necessarily symmetric and we have
\beq
\label{Fbg}
F^{bg}_{\mu\nu}=-C_{[\mu\nu]\sigma}x^\sigma\,.
\eeq
So in the first term in \eqref{L2} which does not have any $C$, both $F_{\mu\nu}$'s are evaluated on perturbations and therefore it can be written as $ff$ up to total derivatives. In the second term with one $C$ only one $F_{\mu\nu}$ could possibly be evaluated on the background so it can be written as $Cxf\partial a$. And there is no restriction on the last term.

\item Let us recall the role of the full nonlinear symmetrization. It was, essentially, an $x$-dependent field redefinition from $a^\mu$ to $\ba^\mu$, so as to cancel the above non-gauge invariant terms of the forms $Cxf\partial a$ and $CCxx\partial a\partial a$. The field redefinition is an identity at $x=0$; therefore, the first term of \eqref{L2}, which can be written as $ff$, remains the same in $\mathcal{L}_\ba^{(2)}$. However, the $x$ dependence of the field redefinition has the byproduct of generating $C\baf\ba$ and $CC\ba\ba$. These two terms, upon introduction of the longitudinal mode \eqref{dpi} and diagonalization, give 
\beq
\label{Lpi2}
\mathcal{L}_\pi^{(2)}\sim CC \partial\pi\partial\pi\,.
\eeq

\item Yet there is a shortcut to obtain the leading part of $\mathcal{L}_\ba^{(2)}$ and therefore $\mathcal{L}_\pi^{(2)}$ from the quadratic Lagrangian $\mathcal{L}_a^{(2)}$ \eqref{L2} without having to find the explicit field redefinition to $\ba^\mu$. We already saw that the leading $ff$ terms in $\mathcal{L}_a^{(2)}$and $\mathcal{L}_\ba^{(2)}$ match each other. To find the rest, derive the equations of motion for $a_\mu$ and take its divergence
\beq
\partial_\mu\frac{\delta\mathcal{L}_a^{(2)}}{\delta a_\mu}=0\,;
\eeq
this automatically projects out the gauge invariant part which describes the transverse modes. The result looks like
\beq
\label{CfCCa}
C\partial f+CC\partial a+\mathcal{O}(x)=0\,.
\eeq
The first two terms are exactly what would have been obtained from the $C\baf \ba$ and $CC\ba\ba$ terms in $\mathcal{L}_\ba^{(2)}$. The $\mathcal{O}(x)$ terms, which vanish at $x=0$, typically describe a ghost that emerges away from $x=0$. This is an artifact of the incomplete alignment of field and coordinate bases. Neglecting them, from \eqref{CfCCa} one can reconstruct $\mathcal{L}_\ba$ and obtain $\mathcal{L}_\pi^{(2)}$.

\item To see why higher order terms in \eqref{phixx} are negligible, suppose we included the cubic correction $\delta\phi_{bg}^\mu = D^\mu_{\alpha\beta\lambda}x^\alpha x^\beta x^\lambda$. Up to second order in $x$ it would contribute to $\mathcal{L}_a^{(2)}$ by at most one insertion of nonlinearities
\beq
Dxx\partial a\partial a\,.
\eeq
However, as argued after \eqref{Fbg} these terms with a single insertion of nonlinearities can be recast in the form $Dxxf\partial a$, which in terms of $\ba$ corresponds to $Dx\baf\ba$ terms. This is evidently subleading compared to previously considered terms since, after diagonalization, it yields the kinetic term $DDxx\partial\pi\partial\pi$ for the longitudinal mode.

\item Finally, curvature effects are encoded in $h^{bg}_{\mu\nu}$ and therefore contribute through $\mathcal{L}_{hA}$. As before we extract terms that survive $x\to 0$ limit. It turns out that only two pieces in $\mathcal{L}_{hA}$ are relevant, namely, 
\beq
\label{hX}
h^{bg}_{\mu\nu}X_2^{\mu\nu}\,,\qquad\text{and}\qquad h^{bg}_{\mu\nu}X_3^{\mu\nu}\,,
\eeq
where one $A_\mu$ in $X_3^{\mu\nu}$ is put on the linearized background. The other terms in $\mathcal{L}_{hA}$ fall into two categories: (a) terms that contain at least one $F_{\mu\nu}$ and (b) terms with higher powers of $h_{\mu\nu}$. In the first case, after symmetrization of the linear part of $\phi_{bg}^\mu$, $F_{\mu\nu}$ in these terms can be evaluated either on the nonlinear part of the background which gives extra suppression in $x$ or should be evaluated on the perturbations to give $f\partial a$-type terms multiplied by $h^{bg}_{\mu\nu}\sim \mathcal{O}(xx)$, which are subleading. Similarly, the type (b) terms are suppressed by extra powers of $x$ after evaluating all $h_{\mu\nu}$ on the background $h^{bg}_{\mu\nu}$.\footnote{The contribution of curvature ($R$) to dynamical mixing with gravity is of the form $\mathcal{L}_{h\pi}^{(2)}\sim Rxxh\partial\partial\pi$ and the only way it may survive $x\to 0$ limit is to give $\mathcal{L}_{h\pi}^{(2)}\sim R h\pi$, which is negligible at large momenta.}

\item As explicitly derived in Appendix \ref{app:quad}, one can use integration by parts and the symmetry properties of $X_n^{\mu\nu}$ to transform \eqref{hX} into the form
\beq
\mathcal{L}^{(2)}\sim R(aa+xaf+xxff)\,,
\eeq
where $R$ stands for Riemann tensor. The leading contribution comes from the first term $Raa$ which using \eqref{dpi} gives
\beq
\mathcal{L}_{\pi}^{(2)}\sim R\partial\pi\partial\pi\,.
\eeq

\end{enumerate}
\subsection{Summary}

There are, therefore, three types of contributions to the kinetic term of the scalar mode: mixing with gravity, nonlinearities of scalar fields $C$, and curvature $R$. After restoring mass scales the final result schematically reads
\beq
\label{Lpi22}
\mathcal{L}_\pi^{(2)}\sim \mpl^2m^4 \partial\pi\partial\pi+\mpl^2m^2 CC\partial\pi\partial\pi+\mpl^2m^2R \partial\pi\partial\pi\,.
\eeq
While around Minkowski or other approximately linear backgrounds mixing with gravity (the first term) plays the main role, once the scale of nonlinearities or curvature becomes comparable to the graviton mass the latter contributions start to dominate.

Therefore, given the strict upper bounds on the value of the graviton mass, even modest deviations from flat space can drastically change the behavior of the scalar mode. For instance, for some choices of parameters of dRGT, the scalar mode becomes a ghost in the self-accelerated solution \cite{giga4} (of course, this is not the Boulware-Deser ghost since there are still five degrees of freedom). 

Finally, denoting the length scale of nonlinearities or curvature by $l$, \eqref{Lpi22} implies that the new cutoff of the theory around those backgrounds is given by
\beq
\Lambda_l = \left(\frac{\mpl m}{l}\right)^{1/3}\,,
\eeq
which can be much larger than $\Lambda_3=(\mpl m^2)^{1/3}$.

\comment{To summarize, The leading quadratic Lagrangian for vector modes can be found by linearizing the backgrounds which after restoring mass scales is of the form
\beq
\mathcal{L}_a^{(2)}\sim \mpl^2m^2ff\,,
\eeq
where the coefficient and the matrices that contract indices of $f_{\mu\nu}$'s depend on the specific background. Typically they are of the order of $\partial_\mu\phi_{bg}^\nu$ which is $\delta^\nu_\mu$ on Minkowski background but can deviate significantly around macroscopic solutions (for instance through Vainshtein mechanism \cite{stars}). See \cite{giga4,KoyamaSA} for eThese deviations can significantly change the behavior of the vector modes 
}

\section*{Acknowledgments}

I am grateful to G. D'Amico, S. Dubovsky, G. Gabadadze, and A. Gruzinov, through several discussions with whom the proof materialized, and L. Berezhiani and R. Rosen for their useful comments on the manuscript. This work was supported by the Mark Leslie graduate award at NYU.

\appendix
\section{\label{app:fullder}Total Derivatives and dRGT}

It was shown in \cite{Nicolis} that in $d$ dimension for each $n\leq d$ there is a unique $n$th order total derivative scalar made out of $\Pi_{\mu\nu}=\partial_\mu\partial_\nu\pi$. Using the Levi-Civita symbol, this scalar combination can be nicely represented as \cite{giga4}
\beq
\label{Lder}
\mathcal{L}^{(n)}_{der}(\Pi)=\vep^{\alpha_1\cdots\alpha_{d-n}\mu_1\cdots\mu_n}{\vep_{\alpha_1\cdots\alpha_{d-n}}}^{\nu_1\cdots\nu_n}\Pi_{\mu_1\nu_1}\cdots\Pi_{\mu_n\nu_n}\,.
\eeq
Since $\mathcal{L}_{der}^{(n)}$ is a total derivative, it can also be written as
\beq
\mathcal{L}_{der}^{(n)}(\Pi)=\partial_\mu\left(\partial_\nu\pi X_{n-1}^{\mu\nu}(\Pi)\right)\,,
\eeq 
where 
\beq
\label{XPi}
X^{\mu\nu}_{n-1}(\Pi)=\frac{1}{n}\frac{\partial \mathcal{L}_{der}^{(n)}}{\partial \Pi_{\mu\nu}}= \vep^{\alpha_1\cdots\alpha_{d-n}\mu\mu_1\cdots\mu_{n-1}}{\vep_{\alpha_1\cdots\alpha_{d-n}}}^{\nu\nu_1\cdots\nu_{n-1}}\Pi_{\mu_1\nu_1}\cdots\Pi_{\mu_{n-1}\nu_{n-1}}\,,
\eeq
and by definition it must be conserved and symmetric ($\partial_\mu X_{n-1}^{\mu\nu}=0$). 

Our goal here is to show that the Lagrangian 
\beq
\label{LderS}
\mathcal{L}_{der}^{(n)}(S_{\mu\nu}/2)=\frac{1}{2^n} \vep^{\alpha_1\cdots\alpha_{d-n}\mu_1\cdots\mu_n}{\vep_{\alpha_1\cdots\alpha_{d-n}}}^{\nu_1\cdots\nu_n}S_{\mu_1\nu_1}\cdots S_{\mu_n\nu_n}\,,
\eeq
which gives $L^{(n)}_{der}(\Pi)$ upon substitution $A_\mu=\partial_\mu\pi$, contains two antisymmetric $F_{\mu\nu}$ up to total derivative terms. Recall that showing the existence of only one $F_{\mu\nu}$ is sufficient, since its antisymmetry guarantees that they come in pairs.

Let us write $S_{\mu\nu}$ as
\beq
S_{\mu\nu}=2\partial_\mu A_\nu-F_{\mu\nu}\,,
\eeq
and substitute it in \eqref{LderS}. The term that contains no $F_{\mu\nu}$ will be
\beq
&2^n\vep^{\alpha_1\cdots\alpha_{d-n}\mu_1\cdots\mu_n}
{\vep_{\alpha_1\cdots\alpha_{d-n}}}^{\nu_1\cdots\nu_n}
\partial_{\mu_1}A_{\nu_1}\cdots \partial_{\mu_n}A_{\nu_n}&\nonumber\\
&=2^n\partial_{\mu_1}\left(\vep^{\alpha_1\cdots\alpha_{d-n}\mu_1\cdots\mu_n}{\vep_{\alpha_1\cdots\alpha_{d-n}}}^{\nu_1\cdots\nu_n} A_{\nu_1}\cdots \partial_{\mu_n}A_{\nu_n}\right)\,,&
\eeq
a total derivative. So we must pick at least one $F_{\mu\nu}$. To see explicitly how it will be paired, rewrite all the rest of $\partial_\mu A_\nu$'s back in terms of $S_{\mu\nu}$ and $F_{\mu\nu}$. Then observe that the product of two Levi-Civita symbols is symmetric under the exchange of all $\{\mu_i\}$ with all $\{\nu_i\}$, and therefore only terms with an even number of $F_{\mu\nu}$'s survive.

\section{\label{app:hpi}Coupling to Gravity}

Consider a generally covariant action
\beq
S=-\frac{1}{4}\mpl^2m^2\int d^4x\sqrt{-g}~V(H^\sigma_\nu)\,,
\eeq
where $V$ is a function of trace of different products of $H^\sigma_\nu=g^{\sigma\mu}H_{\mu\nu}$, and 
\beq
H_{\mu\nu}=g_{\mu\nu}-\partial_\mu\phi^\sigma\partial_\nu\phi^\rho \eta_{\sigma\rho}\,.
\eeq
Our objective is to find the leading coupling between $h_{\mu\nu}=g_{\mu\nu}-\eta_{\mu\nu}$ and the scalar St\"{u}eckelberg $\pi$, when $V$ is chosen to be dRGT. That is, when pure $\pi$ interactions are zero or total derivatives. As a reminder, $\pi$ and $V^\mu$ are defined as follows
\beq
\label{A2}
&\phi^\mu=x^\mu-A^\mu\,,&\\
\label{pi}
&A_\mu=\eta_{\mu\nu}A^\nu=V_\mu+\partial_\mu\pi\,.&
\eeq

The idea is to use the general covariance of the original theory formulated in terms of four scalar fields to establish a relation between different interaction terms in the action. Since $\phi^\mu$ are scalars, under an infinitesimal coordinate transformation (or diffeomorphism) $x'^\mu=x^\mu+\xi^\mu$, we have
\beq
\phi^\mu=x^\mu-A^\mu\quad \to \quad x'^\mu-A^\mu(x')=x^\mu-(A^\mu(x')-\xi^\mu)\,.
\eeq
This shift of $A^\mu$ can be attributed to the vector component $V^\mu$ so that $\pi$ would transform as a scalar
\beq
V_\mu(x)&\to& V_\mu(x')-\xi_\mu\,,\\
\pi(x)&\to& \pi(x')\,.
\eeq
In addition, the metric perturbation $h_{\mu\nu}$ gets shifted under the above coordinate transformation
\beq
h_{\mu\nu}(x)\to h_{\mu\nu}(x')+\partial_{(\mu}\xi_{\nu)}\,.
\eeq

So, although the original action was trivially diffeomorphism invariant, since we are expanding it around a noninvariant background ($g_{\mu\nu}=\eta_{\mu\nu}$ and $\phi^\mu=x^\mu$), the symmetry is nonlinearly realized by $V_\mu$ and $h_{\mu\nu}$. Therefore, the effective action for $h_{\mu\nu}$, $V_\mu$, and $\pi$, which is intrinsically diffeomorphism invariant, fulfills this through a cancellation of the following transformations:
\beq
\label{hxi}
h_{\mu\nu}&\to& h_{\mu\nu}+\partial_{(\mu}\xi_{\nu)}\,,\\
\label{Axi}
V_\mu&\to& V_\mu-\xi_\mu\,,\\
\label{dxi}
\partial_\mu&\to& \partial_\mu+\partial_\mu\xi^\nu\partial_\nu\,,
\eeq
and their higher order corrections which we do not need here. As always, there is also a change of variables of integration from $x^\mu$ to $x'^\mu$. 

Let us apply this to an interaction which is linear in $h_{\mu\nu}$ and of $n$th order in $\Pi_{\mu\nu}$, which we write as $h_{\mu\nu}(\Pi^n)^{\mu\nu}$. Under \eqref{hxi} it changes by 
\beq
\label{xiPin}
\partial_{(\mu}\xi_{\nu)}(\Pi^n)^{\mu\nu}+\mathcal{O}(h_{\mu\nu})\,.
\eeq
This change can only be compensated by the change of the following terms in the effective Lagrangian:
\beq
\partial_{(\mu}V_{\nu)}(\Pi^n)^{\mu\nu}\,\qquad \text{and}\qquad  (\Pi^n)\,,
\eeq
respectively under the transformations \eqref{Axi} and \eqref{dxi}. However, both of these terms are absent in dRGT up to total derivatives. Therefore, \eqref{xiPin} should also vanish up to a total derivative, or
\beq
\partial_\mu(\Pi^n)^{\mu\nu}=0\,.
\eeq
This equation should be familiar from Appendix \ref{app:fullder}, given the conserved $(\Pi^n)^{\mu\nu}$ one can construct an $n+1$st order total derivative
\beq
\mathcal{L}_{der}^{n+1}=\partial_\mu\left(\partial_\nu\pi (\Pi^n)^{\mu\nu}\right)\,.
\eeq
Therefore, we can identify $(\Pi^n)^{\mu\nu}$ with $X_n^{\mu\nu}$ defined in \eqref{XPi} and conclude that the interactions of one $h_{\mu\nu}$ and $\pi$ in $4d$ dRGT are of the form
\beq
\mathcal{L}_{h\pi}= h_{\mu\nu}\sum_{n=1}^{3} a_nX_n^{\mu\nu}(\Pi)+\mathcal{O}(h_{\mu\nu}^2)\,.
\eeq

\section{\label{app:quad}Quadratic Effective Action}

To calculate the effective action for perturbations around a nonlinear background, one often needs to replace $\partial_\mu A^{bg}_\nu$ with a symmetric but not necessarily integrable matrix $(d\tA^{bg})_{\mu\nu}$. Moreover the perturbations $\partial_\mu a_\nu$ should be replaced by $\partial_\mu \ta_\nu+C^\sigma_{\mu\nu}\ta_\sigma$. The resulting quadratic Lagrangian would in general be
\beq
\label{L2B}
\mathcal{L}^{(2)}_\ta=B^{\mu\nu\rho\sigma}(\partial_\mu \ta_\nu+C^\alpha_{\mu\nu}\ta_\alpha)(\partial_\rho \ta_\sigma+C^\beta_{\rho\sigma}\ta_\beta)\,,
\eeq
where $B^{\mu\nu\rho\sigma}$ depends on the background fields. 

Our aim is to show that in dRGT, $\mathcal{L}^{(2)}_\ta$ consist only of $\tf\tf$, $\tf\ta$, and $\ta\ta$. When the $C\ta$ term is chosen from both parentheses in \eqref{L2B}, this is automatic. When one $\partial\ta$ and one $C\ta$ are chosen, we have
\beq
\mathcal{L}_{\partial \ta \ta}^{(2)}=(BC)^{\mu\nu\rho}\partial_\mu\ta_\nu \ta_\rho\,.
\eeq
This may not generically look like $\tf \ta$ but can be transformed into that by integration by parts. Writing $BC$ in terms of symmetric and antisymmetric parts 
\beq
(BC)^{\mu\nu\rho}=\frac{1}{2}\left[(BC)^{[\mu\nu]\rho}+(BC)^{(\mu\nu)\rho}\right]
\eeq
and denoting equality up to total derivatives by $\cong$, we have
\beq
\label{ada}
\mathcal{L}_{\partial \ta \ta}^{(2)}&=&\frac{1}{2}(BC)^{[\mu\nu]\rho}\partial_\mu\ta_\nu \ta_\rho + \frac{1}{2}(BC)^{(\mu\nu)\rho}\partial_\mu\ta_\nu \ta_\rho\\
&\cong&\frac{1}{4}(BC)^{[\mu\nu]\rho}\tf_{\mu\nu} \ta_\rho-\frac{1}{2}\partial_\mu(BC)^{(\mu\nu)\rho}\ta_\nu \ta_\rho-\frac{1}{2}(BC)^{(\mu\nu)\rho}\ta_\nu \partial_\mu\ta_\rho \nonumber\\
&=&\frac{1}{4}(BC)^{[\mu\nu]\rho}\tf_{\mu\nu} \ta_\rho-\frac{1}{2}\partial_\mu(BC)^{(\mu\nu)\rho}\ta_\nu \ta_\rho-\frac{1}{2}(BC)^{(\mu\nu)\rho}\ta_\nu (\tf_{\mu\rho}+\partial_\rho\ta_\mu)\nonumber\\
&=& \frac{1}{4}(BC)^{[\mu\nu]\rho}\tf_{\mu\nu} \ta_\rho-\frac{1}{2}\partial_\mu(BC)^{(\mu\nu)\rho}\ta_\nu \ta_\rho-\frac{1}{2}(BC)^{(\mu\nu)\rho}\ta_\nu \tf_{\mu\rho} - \frac{1}{4}(BC)^{(\mu\nu)\rho}\partial_\rho(\ta_\nu \ta_\mu)\nonumber\\
&\cong& \frac{1}{4}(BC)^{[\mu\nu]\rho}\tf_{\mu\nu} \ta_\rho-\frac{1}{2}\partial_\mu(BC)^{(\mu\nu)\rho}\ta_\nu \ta_\rho-\frac{1}{2}(BC)^{(\mu\nu)\rho}\ta_\nu \tf_{\mu\rho} + \frac{1}{4}\partial_\rho(BC)^{(\mu\nu)\rho}\ta_\nu \ta_\mu\nonumber
\eeq
which is of the desired form.

So we only need to check $\partial\ta\partial\ta$ terms. When the indices of $\partial_\mu \ta_\nu$ are antisymmetrized, as is the case for the contribution of all higher than quartic terms in $\mathcal{L}_A$, we also get the expected form $\tf\tf$. It remains to check the three $\mathcal{L}_{der}^{(n)}(S_{\mu\nu}/2)$ terms. On the symmetric background [$S^{bg}_{\mu\nu}=2(d\tA^{bg})_{\mu\nu}$], they give
\beq
\label{Lder2}
\mathcal{L}_{der}^{(n,2)}=\frac{n(n-1)}{8}\vep^{\alpha_1\cdots\alpha_{d-n}\mu_1\cdots\mu_n}{\vep_{\alpha_1\cdots\alpha_{d-n}}}^{\nu_1\cdots\nu_n}(d\tA^{bg})_{\mu_3\nu_3}\cdots(d\tA^{bg})_{\mu_n\nu_n}\ts_{\mu_1\nu_1}\ts_{\mu_2\nu_2}\,.
\eeq
After writing $\ts_{\mu\nu}=2\partial_\mu\ta_\nu-\tf_{\mu\nu}$, three types of terms arise: $\partial_{\mu_1}\ta_{\nu_1}\partial_{\mu_2}\ta_{\nu_2}$, in which the $\partial_{\mu_1}$ derivative can be integrated by parts to act on the background-dependent coefficient. It therefore becomes $\ta_{\nu_1}\partial_{\mu_2}\ta_{\nu_2}$, which is harmless according to \eqref{ada}. The mixed term $\partial_{\mu_1}\ta_{\nu_1}\tf_{\mu_2\nu_2}$ gives $\tf_{\mu_1\nu_1}\tf_{\mu_2\nu_2}/2$ because of symmetry reasons. And the last possibility is to choose both $\tf_{\mu\nu}$'s, which is automatically of the desired form.

The introduction of nonflat background metric $\bg_{\mu\nu}$ does not change the above conclusion since it does not change the symmetry of indices and can be viewed as a part of the background-dependent coefficients.

Finally, consider the case encountered in Sec. \ref{scalar}, where we want to find the leading quadratic Lagrangian without fully aligning field and coordinate bases. In this case there is no $x$-dependent field redefinition and the curvature effects are encoded in $h^{bg}_{\mu\nu}$. One should therefore study $h_{\mu\nu}X_n^{\mu\nu}(S_{\mu\nu}/2)$ terms in $\mathcal{L}_{hA}$, when $h_{\mu\nu}$ and all $S_{\mu\nu}$'s in $X_n^{\mu\nu}$ except two are evaluated on the background
\beq
\label{LhX2}
\mathcal{L}_{hX_n}^{(2)}=\frac{n(n-1)}{2^{n+1}}h^{bg}_{\mu\nu}\vep^{\alpha_1\cdots\alpha_{d-n}\mu\mu_1\cdots\mu_{n-1}}{\vep_{\alpha_1\cdots\alpha_{d-n}}}^{\nu\nu_1\cdots\nu_{n-1}}S^{bg}_{\mu_3\nu_3}\cdots S^{bg}_{\mu_{n-1}\nu_{n-1}} s_{\mu_1\nu_1} s_{\mu_2\nu_2}\,.
\eeq
This would look very similar to \eqref{Lder2} if we replaced $h^{bg}_{\mu\nu}$ by $S^{bg}_{\mu\nu}$. Nevertheless the argument following \eqref{Lder2} used only the symmetry of the background fields and, since $h^{bg}_{\mu\nu}$ has the same symmetry as $S^{bg}_{\mu\nu}$, it applies to this case as well. Thus \eqref{LhX2} is reducible to $ff$, $fa$, and $aa$ terms. In the case of interest, where $h^{bg}_{\mu\nu}\sim R xx$, one can ignore nonlinearities in $A^{bg}_\mu$ and take $S^{bg}_{\mu\nu}$ to be constant. \eqref{LhX2} can then be manipulated to give
\beq
\mathcal{L}_{hX_n}^{(2)}&=&-\frac{n(n-1)}{2^{n+1}}\vep^{\alpha_1\cdots\alpha_{d-n}\mu\mu_1\cdots\mu_{n-1}}{\vep_{\alpha_1\cdots\alpha_{d-n}}}^{\nu\nu_1\cdots\nu_{n-1}}S^{bg}_{\mu_3\nu_3}\cdots S^{bg}_{\mu_{n-1}\nu_{n-1}}\nonumber\\
&&\times \left(4 a_{\mu_2} a_{\nu_2}\partial_{\mu_1}\partial_{\nu_1}
  +3 f_{\nu_1\nu_2}a_{\mu_2}\partial_{\mu_1}+2 f_{\mu_2\nu_2}a_{\nu_1}\partial_{\mu_1}+\frac{1}{2}f_{\mu_1\mu_2}f_{\nu_1\nu_2}\right)h^{bg}_{\mu\nu} \,.
\eeq


\end{document}